\documentclass[fleqn,usenatbib]{mnras}

\usepackage{newtxtext,newtxmath}

\usepackage[T1]{fontenc}

\DeclareRobustCommand{\VAN}[3]{#2}
\let\VANthebibliography\thebibliography
\def\thebibliography{\DeclareRobustCommand{\VAN}[3]{##3}\VANthebibliography}

\usepackage{graphicx}	
\usepackage{amsmath}	
\usepackage{pgffor}
\usepackage{xtab,booktabs}
\usepackage{longtable}

\newcommand{\DsDl}{$\mathcal{D}_{S}/\mathcal{D}_{L}\,$}
\newcommand{\DtoZ}{$\mathcal{D}/Z\,$}
\newcommand{\Dtot}{$\mathcal{D}_{\text{tot}}\,$}
\newcommand{\fmol}{$f_{\text{H}_{2}}\,$}
\newcommand{\Vanilla}{\emph{Vanilla}}
\newcommand{\NoISRF}{\emph{NoISRF}}
\newcommand{\ESFB}{\emph{ESFB}}
\newcommand{\NoGSD}{\emph{NoGSD}}
\newcommand{\NoSh}{\emph{NoSh}}
\newcommand{\NoPr}{\emph{NoPr}}
\newcommand{\Early}{\emph{Z0.1ICs}}
\newcommand{\Late}{\emph{Z0.5ICs}}

\defcitealias{1977ApJ...217..425M}{MRN}
\defcitealias{2017MNRAS.467..699H}{HH17}
\defcitealias{2018MNRAS.474.1545C}{CH18}
\defcitealias{2019MNRAS.482.2555H}{HA19}
\defcitealias{2020MNRAS.491.3844A}{AHN20}
\defcitealias{2022arXiv220205243R}{RNH22}

\title[Co-evolution of H$_2$ and the grain size distribution]{The co-evolution of molecular hydrogen and the grain size distribution in an isolated galaxy}

\author[L. Romano et al.]{
Leonard E. C. Romano$^{1,2}$\thanks{E-mail: leonard.romano@tum.de},
Kentaro Nagamine$^{2,3,4}$ and
Hiroyuki Hirashita$^{5}$
\\
$^{1}$Physik-Department, Technische Universität München, James-Franck-Straße, 85748 Garching, Germany\\
$^{2}$Theoretical Astrophysics, Department of Earth and Space Science, Osaka University, 1-1 Machikaneyama, Toyonaka, Osaka 560-0043, Japan\\
$^{3}$Kavli IPMU (WPI), The University of Tokyo, 5-1-5 Kashiwanoha, Kashiwa, Chiba 277-8583, Japan\\
$^{4}$Department of Physics and Astronomy, University of Nevada, Las Vegas, 4505 S. Maryland Pkwy, Las Vegas, NV 89154-4002, USA\\
$^{5}$Institute of Astronomy and Astrophysics, Academia Sinica, Astronomy-Mathematics Building, AS/NTU, No. 1, Section 4, Roosevelt Road, Taipei 10617, Taiwan \\
}

\date{Accepted XXX. Received YYY; in original form ZZZ}

\pubyear{2021}

\begin{document}
\label{firstpage}
\pagerange{\pageref{firstpage}--\pageref{lastpage}}
\maketitle

\begin{abstract}
Understanding the evolution of dust and molecular hydrogen (H$_2$) is a critical aspect of galaxy evolution, as they affect star formation and the spectral energy distribution of galaxies. 
We use the $N$-body/smoothed-particle-hydrodynamics code {\sc Gadget4-Osaka} to compute the evolution of dust and H$_2$ in a suite of numerical simulations of an isolated Milky-Way-like galaxy. 
The evolution of the full grain size distribution (GSD) is solved by sampling the grain size on a logarithmically spaced grid with 30 bins. The evolution of a primordial chemistry network with twelve species is solved consistently with the hydrodynamic evolution of the system, including star formation, metal and energy ejections from stars into the interstellar medium through supernova feedback and stellar winds. 
The formation model for H$_2$ considers the GSD and photo-dissociation through the UV radiation of young stars. We identify the processes needed for producing a sizeable amount of H$_2$, verify that the resulting star formation law in the later stages of galaxy evolution is consistent with observations of local spirals, and show that our model manages to produce a galactic molecular gas fraction in line with observations of Milky-Way-like galaxies. We stress the importance of the co-evolution of the GSD and H$_2$, as models assuming a fixed MRN shape for the GSD overestimate the production of H$_2$ in regimes where the dust abundance is dominated by large grains and underestimate it in the regime where the dust is dominated by small grains, both of which are realized in simulations of dust evolution.

\end{abstract}




\section{Introduction}
\label{sec:intro}

Galaxy evolution is greatly dependent on star formation. Given that star formation predominantly occurs in molecular clouds, understanding their formation and evolution is an important step towards a comprehensive understanding of the relation between star formation and galaxy evolution. While H$_2$ is the most abundant constituent of these clouds, it is hard
to detect due to its weak emission. Instead typically the H$_2$ mass is traced by carbon monoxide (CO) \citep[e.g.][]{1975ApJ...199L.105S, 1986ApJ...309..326D, 2013ARA&A..51..207B, 2017MNRAS.468L.103H} or by [\ion{C}{ii}] \citep[e.g.][]{1997ApJ...483..200M, 1998ApJ...498..735P, 1999RvMP...71..173H}. 
While the conversion factors between these species and the H$_2$ abundance are motivated by empirical relations in the local Universe, the detection of heavy elements and carbon-rich molecules in collapsing metal-enriched clouds suggests that
they consist of a CO core which is surrounded by a \ion{C}{ii} envelope hosting H$_2$ \citep[e.g.][]{2006A&A...451..917R, 2010ApJ...716.1191W, 2012ApJ...746...69G, 2013ARA&A..51..207B, 2018MNRAS.481.1976Z, 2020A&A...643A.141M, 2020ApJ...903..142G}. The current local estimate of the H$_2$ mass density in the local Universe based on these techniques is $\rho_{\text{H}_2} \approx 6.8 \times 10^6 \, \text{M}_{\sun}\, \text{Mpc}^{-3}\,$ amounting to roughly 20 percent of the total cold gas mass \citep{2018MNRAS.476..875C, 2021MNRAS.501..411F}.
The column densities of molecular hydrogen and the star formation rate in the local Universe have been shown to correlate \citep[e.g.][]{1998ApJ...498..541K, 2008AJ....136.2846B, 2012ApJ...745..183R, 2013AJ....146...19L}. This relation gives important hints to how quickly molecular gas is depleted and converted into stars, helping to constrain theoretical star formation models \citep[e.g.][]{2016ApJ...826..200S}. However, the universal applicability of this relation at all epochs is still debated. Two important channels of keeping a net H$_2$ abundance in the metal-enriched interstellar medium (ISM) is the formation on grain surfaces \citep{1963ApJ...138..393G} and the shielding from the interstellar ultra-violet (UV) radiation due to dust extinction,
both of which depend on the total grain surface area determined by the dust abundance and the grain size distribution (GSD) \citep[e.g.][henceforth HH17]{2011ApJ...735...44Y, 2017MNRAS.467..699H}. It is thus important to study the evolution of molecular gas alongside the evolution of the GSD. 

Both, the abundances of H$_2$ and dust grains evolve as star formation progresses. Before the onset of star formation, the gas in the ISM is devoid of metals and grains; therefore, 
H$_2$ production is dominated by catalysis of intermediate species with H \citep{1998A&A...335..403G}; H$^{-}$ catalysis (H $+$ e$^{-}$ $\rightarrow$ H$^{-} + \gamma$ and H + H$^{-} \rightarrow$ H$_2$ + e$^{-}$) and H$_2^{+}$ catalysis (H $+$ H$^{+} \rightarrow$ H$_2^{+} + \gamma$ followed by H$_2^{+} + $ H $\rightarrow$ H$_2 + $ H$^{+}$). In sufficiently dense, neutral media three-body reactions (3H $\rightarrow$ H$_2$ + H) can also contribute. After the onset of star formation, strong UV
radiation from young stars can lead to photodissociation of the molecules, unless they are condensed in clouds, which are sufficiently dense for shielding to be effective \citep[e.g.][]{1996ApJ...468..269D, 2011MNRAS.418..838W, 2014ApJ...790...10S}. Shortly after the onset of star formation, supernova (SN) explosions can dissociate molecules in the post-shock region \citep{1977ApJ...216..713K} and heat up the ISM, creating a hostile environment for the survival of molecular species. At the same time, dust grains condense from the metals in the ejecta of SN explosions \citep[e.g.][]{1989ApJ...344..325K, 2007MNRAS.378..973B, 2007ApJ...666..955N} and the stellar winds of asymptotic giant branch (AGB) stars \citep[e.g.][]{2006A&A...447..553F, 2014MNRAS.439..977V, 2017MNRAS.467.4431D}, marking the onset of dust evolution as described by the comprehensive model by \citet{2013MNRAS.432..637A}. 

The dust grains created by stellar sources are typically large with sizes around $a \sim 0.1\, \micron$ \citep{2012ApJ...745..159Y}, which are subsequently subject to processing. In the warm, diffuse ISM the large grains can efficiently shatter, leaving behind small fragments \citep[e.g.][]{2009MNRAS.394.1061H}. Once these small fragments are transported to the cold and dense medium, they can grow through accretion of gas phase metals and coagulate to form larger grains \citep[e.g.][]{2012MNRAS.422.1263H} evolving the GSD towards a power law shape similar to the one observed by \citet[][henceforth MRN]{1977ApJ...217..425M}. The dust mass in both large and small grains is gradually reduced by sputtering in SN shocks and in hot gas with temperature $\gtrsim 10^6\,\mathrm{K}\,$.
In this picture, the GSD is initially dominated by large grains, which over time get processed, leading to the gradual enrichment with small grains.
Once the ISM has been sufficiently enriched with dust grains, H$_2$ formation on grain surfaces might become the dominant production mechanism, depending on the efficiency of photoelectric heating and recombination cooling \citep[e.g.][]{1994ApJ...427..822B}.

While this basic picture is widely recognized \citep[e.g.][henceforth CH18]{2011ApJ...728...88G, 2021arXiv211113701M, 2018MNRAS.474.1545C}, the relative importance of the H$_2$ formation processes at different stages of galaxy evolution is still a matter of active research \citep[see e.g.][]{2015MNRAS.452.3815L, 2017MNRAS.471.4128P, 2017MNRAS.467..115D, 2018ApJS..238...33D, 2019MNRAS.485.4817D}.

Dust evolution and its impact on galaxy evolution has been studied in a number of numerical simulations. Early simulations usually did not resolve the GSD and instead followed the total dust mass assuming a fixed GSD \citep[e.g.][]{2014MNRAS.439.3073Y, 2015MNRAS.449.1625B, 2016ApJ...831..147Z, 2016MNRAS.457.3775M, 2017MNRAS.468.1505M}. Later a simplified dust evolution model, where the GSD is divided into two bins referred to as `small' and `large' grains, known as the two-size approximation \citep{2015MNRAS.447.2937H} has been used in simulations of astrophysical dust evolution \citep[e.g.][]{2017MNRAS.466..105A, 2017MNRAS.469..870H, 2018MNRAS.479.2588G, 2021MNRAS.503..511G}. More recently, the evolution of the full GSD has been included in simulations of isolated galaxies \citep[][]{2018MNRAS.478.2851M, 2020MNRAS.491.3844A} and a cosmological box directly \citep{2021MNRAS.507..548L} and with post-processing \citep{2021MNRAS.501.1336H}. 

There have also been a number of simulations in which the evolution of molecular hydrogen was modelled \citep[e.g.][]{2009ApJ...697...55G, 2018MNRAS.474.2884L, 2021arXiv211113701M}, even though only a few have taken into account the dependence on grain size.
\citetalias{2017MNRAS.467..699H} applied the two-size approximation in a one-zone model of clouds with hydrogen column density $N_\text{H} = 10^{21}$ -- $10^{23}\,\text{cm}^{-2}\,$, in order to study the effect of grain size on the evolution on the abundances of H$_2$ and CO in the ISM and the CO-to-H$_2$ conversion factor. They confirmed that grain size has a non-negligible effect on the evolution of these molecules. However, since their results were based on a one-zone model, they could not predict the spatial distribution of the molecules. In order to address this limitation, \citetalias{2018MNRAS.474.1545C} post-processed snapshots of a hydrodynamic simulation of an isolated galaxy with dust evolution based on the two-size approximation from \citet{2017MNRAS.466..105A}, studying the equilibrium abundances of H$_2$ and CO using \citetalias{2017MNRAS.467..699H} as a subgrid model for unresolved dense clouds.
They also find that grain size evolution plays an important role for the synthesis of molecular species in galaxies, especially at low metallicity.

The aim of this study is to further understand the relative importance of the different channels of H$_2$ production and dissociation at different stages of galaxy evolution. To this end we run a suite of hydrodynamic simulations of an isolated galaxy modelling the evolution of the full GSD based on the model presented in a companion paper, \citet[][henceforth RNH22]{2022arXiv220205243R}, and the grain size dependent, non-equilibrium evolution of molecular hydrogen. The model for the GSD developed in \citetalias{2022arXiv220205243R} is based on the model by \citet{2020MNRAS.491.3844A}, but includes diffusion of dust and metals.

In this work, we appreciate the importance of different processes at different stages of the galactic evolution by comparing the evolution of the galactic molecular gas content between runs where the processes are artificially switched on and off, and study how the molecular abundance scales with a number of related observables. Moreover, we confirm that our results are consistent with available observational data and show that a molecular star formation law arises in our model under certain physical conditions.

This paper is organized as follows. In Section \ref{sec:methods}, we describe the numerical simulations and the model for H$_2$ formation on dust grains and photodissociation in the interstellar radiation field (ISRF). In Section \ref{sec:results}, we present the results of our simulations and compare them to observations where possible. In Section \ref{sec:discussion}, we provide a discussion of our results and conclude by giving a summary of our findings.
As in our companion paper \citepalias{2022arXiv220205243R}, we adopt a value of $Z_{\sun} = 0.01295$.
  
\section{Methods}
\label{sec:methods}

\subsection{Hydrodynamic Simulation}\label{sec:simulation}
We simulate the hydrodynamic evolution of an isolated galaxy using the smoothed particle hydrodynamics (SPH) code {\sc Gadget4-Osaka}, which is based on the {\sc Gadget3-Osaka} models \citep[]{2017MNRAS.466..105A,2019MNRAS.484.2632S, 2021ApJ...914...66N} integrated into the {\sc Gadget-4} code \citep{2021MNRAS.506.2871S}. The star formation and production of dust and metals is treated self-consistently with the dynamical evolution of the simulated isolated galaxy. We calculate the evolution of GSD on each gas particle as summarized in Section \ref{sec:dustEvolution}.
We are using the primordial chemistry and cooling library {\sc Grackle-3}\footnote{\label{footer:Grackle3} \href{https://grackle.readthedocs.org/}{https://grackle.readthedocs.org/}}\citep{2017MNRAS.466.2217S}, which solves the non-equilibirum chemistry network for H, He, D and their molecules. We also employ photo-heating, photo-ionization and photo-dissociation due to the UV background radiation (UVB) at $z = 0$ from \citet{2012ApJ...746..125H}. Metal cooling is scaled linearly with the total metallicity. We do not distinguish between metals condensed into dust grains and gas-phase metals.

In order to study the effect the evolution of the GSD has on the formation of molecular hydrogen, we evolve a Milky Way-like galaxy in isolation for $2\,\text{Gyr}\,$, after which we do not expect the molecular abundance to undergo significant changes (see, e.g.\ Fig.~\ref{fig:H2history}). We employ the low resolution initial conditions (ICs) of the AGORA collaboration \citep{2016ApJ...833..202K}, but add a hot ($T_\mathrm{gas} = 10^6\, \text{K}$, where $T_\mathrm{gas}$ is the gas temperature) gaseous halo component with a total mass of $\sim 10^9 \,\text{M}_{\sun}\,$ as described by \citet{2021ApJ...917...12S}. The galactic disk initially hosts a stellar bulge and disk of mass $M_{\star} \sim 4\times 10^{10}\,\text{M}_{\sun}\,$ and a gaseous disk of mass $M_{\text{disk}} \sim 8.5 \times 10^{9}\,\text{M}_{\sun}\,$, which are sampled by $\sim 10^5\,$ collisionless star particles of mass $m_{\star} = 3.437 \times 10^{5}\,\text{M}_{\sun}\,$ and $10^5\,$ gas particles of initial mass $m_{\text{gas, IC}} = 8.593 \times 10^{4}\,\text{M}_{\sun}\,$, respectively. The galaxy is embedded into a dark matter halo with a mass of $M_{200} \sim 10^{12}\,\text{M}_{\sun}\,$, sampled by $10^5$ collisionless DM particles of mass $m_{\text{DM}} = 1.254\times 10^{7}\,\text{M}_{\sun}\,$. Initial elemental abundances are set to their primordial values everywhere and we employ a floor value for metallicity of $Z_{\text{floor}} = 10^{-4}\,Z_{\sun}\,$. Admittedly such a setup is rather unrealistic and unlikely to be observed in nature. Nonetheless, the goal of this study is to understand the evolution of dust and molecules under a wide range of physical conditions (including metal enrichment), and for this purpose such ICs provide a powerful tool for model development. In order to verify the robustness of our results, we discuss the dependence on the ICs in Appendix~\ref{Appendix:ICs}.
In order to avoid numerical fragmentation due to fluctuation in the initial density field, we adiabatically relax the ICs for $500\, \text{Myr}\,$, before evolving these relaxed ICs for another $2\, \text{Gyr}\,$ with enabled subgrid physics. We run six simulations with different models for radiation feedback, self-shielding and H$_2$ production as summarized in Section \ref{sec:runs}. In all of our runs we employ a gravitational softening length of $\epsilon_\text{grav} = 80\, \text{pc}\,$, but allow the SPH smoothing length to get as small as $0.1 \epsilon_\text{grav}$.

\subsection{Dust Evolution Model}\label{sec:dustEvolution}
We use the same model as \citetalias{2022arXiv220205243R} for the evolution of the GSD. The model is an extension to the one used by \citetalias{2020MNRAS.491.3844A} and \citetalias{2019MNRAS.482.2555H}, which is a simplified version of the model by \citet{2013MNRAS.432..637A} suitable for the inclusion in hydrodynamical simulations. \citetalias{2022arXiv220205243R} extended the model by introducing a subgrid model for the turbulent mixing of metals and dust, including thermal sputtering as an additional channel for dust destruction in hot gas and modifiying the subgrid prescription for the unresolved multiphase ISM. We consider dust evolution by stellar dust production, destruction of dust grains by sputtering in SN shocks and hot plasma, dust growth by accretion and coagulation in dense clouds on unresolved scales and shattering by grain-grain collisions in the diffuse ISM. Dust grains are assumed to be electrically neutral, compact spheres that are dynamically coupled to the gas; an assumption that is usually valid at the scales considered \citep{2018MNRAS.478.2851M}. We define the grain mass distribution $\rho_{\text{d}}\left(m\right)\text{d}m$ to be the mass density of grains with mass between $m$ and $m + \text{d}m$. In our simulation, each SPH particle samples the grain mass distribution at 30 grain radii $a$ logarithmically spaced between $3 \times 10^{-4}\,$ and $10\,\micron$, which are related to the grain mass by $m = \left(4\pi/3\right)\,s\,a^3$ where we adopt a value of the bulk density $s = 3.5\,\text{g}\,\text{cm}^{-3}\,$ appropriate for silicates \citep{2001ApJ...548..296W}. For the turbulent diffusion of dust and metals, we choose a diffusion parameter of $C_{\text{d}} = 0.02$, which lead to the best agreement with observed extinction curves in \citetalias{2022arXiv220205243R}.

We refer the interested reader to \citetalias{2020MNRAS.491.3844A} for an outline of the model that is used to evolve the dust abundance for each SPH particle, to \citetalias{2019MNRAS.482.2555H} for the complete set of equations underlying the model and to \citetalias{2022arXiv220205243R} for details of the turbulent diffusion model, the thermal sputtering and the modified prescription for the unresolved ISM.

\subsection{H$_2$ Formation}\label{sec:h2production}

We evolve the H$_2$ abundance of each gas particle consistently with its hydrodynamical and chemical evolution as described in section \ref{sec:simulation}. We extend the work of \citetalias{2018MNRAS.474.1545C}, who studied the effect of grain size on the formation of H$_2$ by post-processing hydrodynamical simulation of the same isolated galaxy which followed the dust evolution using the two-size-approximation \citep{2017MNRAS.466..105A}. We quantify the abundance of molecular hydrogen and other elements using their mass fractions $X_i$, i.e.,
\begin{equation}
    X_{\text{H}_2} = \frac{m_{\text{H}_2}}{m_\text{gas}}.
\end{equation}
The molecular fraction \fmol is related to the mass fraction by
\begin{equation}
    f_{\text{H}_2} = \frac{X_{\text{H}_2}}{X_\text{H}}.
\end{equation}
The formation of H$_2$ in the gas phase is treated by Grackle-3 and will therefore not be described any further. In dense, metal-enriched gas,  formation on grain surfaces plays a dominant role. The process is described by
\begin{equation}
    \frac{dX_{\text{H}_2}}{dt} = 2\, n_{\rm H} \,R_{\text{H}_2}\, X_\text{HI}, 
\end{equation}
where $n_\text{H}$ is the number density of hydrogen and the reaction rate coefficient $R_{\text{H}_2}$ has been computed by \citet{2011ApJ...735...44Y} as follows: 
\begin{equation}\label{eq:formationGSD}
    R_{\text{H}_2} = \frac{1}{2}\, S\, v_\text{th}\, \mu\, m_\text{H} \sum_i \frac{\sigma_i\, \mathcal{D}_i}{m_i}, 
\end{equation}
where $S$ the sticking coefficient, $v_\text{th}$ the thermal velocity, $\mu = 1.4$ is the gas mass per hydrogen, $m_\text{H}$ is the proton mass, $\sigma_i = \pi a_i^2$ is the geometrical cross-section ($i$ indicates the grain radius bin), $m_i = (4\pi/3) s a_i^3$ is the grain mass with size $a_i$, and $\mathcal{D}_i = \rho_{\text{d}}\left(m_{i}\right)\text{d}m_{i}/\rho_{\rm gas}$ is the dust-to-gas ratio of dust grains in the $i$th bin.
For the thermal velocity, we follow the definition by \citet{1978ppim.book.....S}:
\begin{equation}
    v_\text{th} = \sqrt{\frac{8k_\text{B} T_\text{gas}}{\pi m_\text{H}}}, 
\end{equation}
where $k_\text{B}$ is the Boltzmann constant. 

To study the effect of grain size, we also run a simulation with a grain size independent formation rate coefficient. 
For comparison with other studies, we adopt the widely used expression for the formation rate coefficient $R_{\text{H}_2}^\text{NoGSD}$ \citep[see e.g.][]{2009ApJ...697...55G, 2018MNRAS.474.2884L, 2021arXiv211113701M}, instead of the above $R_{\text{H}_2}$:
\begin{equation}\label{eq:formationNoGSD}
    R_{\text{H}_2}^\text{NoGSD} = 3.5 \times 10^{-17} \left(\frac{\mathcal{D}_\text{tot}}{0.01}\right) \,\text{cm}^3\,\text{s}^{-1},
\end{equation}
where \Dtot is the dust-to-gas ratio. We note that for the MRN GSD, the grain-size dependent expression from \citet{2011ApJ...735...44Y} also yields a value of $R_{\text{H}_2} \sim 10^{-17} \left(\mathcal{D}_\text{tot}/0.01\right) \,\text{cm}^3\,\text{s}^{-1}$ for the values of $S$ and $v_\text{th}$ adopted in our subgrid prescription. Formation of H$_2$ by this process is very inefficient unless the densities are high enough, i.e. $n_{\text{H}} \gtrsim 10^3\,\text{cm}^{-3}$. Unfortunately, with our numerical resolution, such dense environments cannot be resolved and we need to rely on a subgrid prescription. To this end, we apply the prescription described in \citetalias{2022arXiv220205243R} in which cold ($T < 10^4\,\text{K}$) gas particles with a density higher than $n_{\text{H}} = 0.1 \,\text{cm}^{-3}\,$ host cold and dense clouds which make up a fraction $f_{\text{cloud}}\left(n_{\text{H}}\right)$ of their mass. Here $f_{\text{cloud}}\left(n_{\text{H}}\right)$ is assumed to increase linearly with density before it saturates with a value of unity, i.e.
\begin{equation}
    f_{\text{cloud}}\left(n_{\text{H}}\right) = \text{min} \left\{1, \alpha\, n_{\text{H}, 0}\right\},
\end{equation}
where we adopt $n_{\text{H}, 0} = \left(n_{\text{H}}/ 1\,\text{cm}^{-3}\right)\,$and $\alpha = 0.12$ was chosen, such that the global fraction of dense gas $f_\text{dense} = \sum_j m_j f_{\text{cloud}, j}$ ($j$ indicates a gas particle in the disk) is roughly 20 percent, roughly corresponding to the typical molecular fraction in MW-like galaxies \citep{2018MNRAS.476..875C}.
The cold clouds have a fixed density and temperature of $n_\text{cloud} = 10^3\,\text{cm}^{-3}\,$, $T_\text{cloud} = 50\,\text{K}\,$, and we fix the sticking coefficient to be $S = 0.3$. H$_2$ formation on dust grains is then only considered for these dense clouds.

\subsection{H$_2$ Dissociation}
\label{sec:dissociation}

One of the main drivers of H$_2$ destruction is the photo-dissociation in the ISRF. Young stars emit strong UV radiation that can dissociate the molecular hydrogen in the surrounding medium via the two-step Solomon-process. The rate at which this can happen is given by
\begin{equation}
    \frac{dX_{\text{H}_2}}{dt} = -\Gamma_{\text{H}_2}X_{\text{H}_2}.
\end{equation}
The rate coefficient $\Gamma_{\text{H}_2}$ depends on the strength of the incident radiation field in the Lyman-Werner (LW) band and is given by \citep{2005MNRAS.356.1529H}
\begin{equation}\label{eq:H2_dissociation}
    \Gamma_{\text{H}_2} = 4.4 \times 10^{-11} \chi S_\text{shield}\,\left[s^{-1}\right], 
\end{equation}
where $\chi$ is the strength of the radiation field in the LW-band in units of the Habing radiation field \citep{1968BAN....19..421H}, and $S_\text{shield}$ takes into account the attenuation through shielding by dust grains and the self-shielding of molecular gas, i.e., $S_\text{shield} = S_{\text{shield, dust}} \times S_{\text{shield, H}_2}$. The shielding factor due to the presence of dust grains is given by
\begin{equation}\label{eq:extinctionGSD}
    S_{\text{shield, dust}} = \exp{\left(-\sum_i \tau_{\text{LW}, i}\right)},
\end{equation}
where the sum runs over grain size bins $i$ and we estimate the optical depth due to grains of size $a_i$ as
\begin{equation}\label{eq:tauLW}
    \tau_{\text{LW}, i} = 0.23 \left(\frac{\mathcal{D}_i}{0.01}\right)\left(\frac{a_i}{0.1\,\micron}\right)^{-1} \left(\frac{N_\text{H}}{10^{20}\, \text{cm}^{-2}}\right).
\end{equation}
In the run where we neglect the grain size dependence, we adopt instead of $\Sigma_i\tau_{\text{LW},i}$
\begin{equation}\label{eq:extinctionNoGSD}
    \tau_{\text{LW}}^{\text{NoGSD}} = 0.23 \left(\frac{\mathcal{D}_\text{tot}}{0.01}\right) \left(\frac{N_\text{H}}{10^{20}\,\text{cm}^{-2}}\right),
\end{equation}
which corresponds to setting $a = 0.1\, \micron\,$ in eq. (\ref{eq:tauLW}).

For the self shielding we employ the formula from \citet{2011MNRAS.418..838W}:
\begin{equation}\label{eq:H2_self_shielding}
    S_{\text{shield, H}_2} = \frac{0.965}{\left(1 + \frac{x}{b_5}\right)^{1.1}} + \frac{0.035}{\sqrt{1 + x}} \exp{\left[- 8.5 \times 10^{-4} \sqrt{1 + x}\right]}, 
\end{equation}
where $x = N_{\text{H}_2} / 5 \times 10^{14}\,\text{cm}^{-2}$,  and $b_5 = \sqrt{2} \sigma_\text{v}$ is the doppler broadening parameter with the velocity dispersion $\sigma_\text{v}$ measured in km s$^{-1}$, which we approximate by the local sound speed. Column densities are estimated by multiplying the density with the local Sobolev-like length scale $L_{\text{sob}} = \rho / \left|\nabla \rho\right|$, which according to \citet{2011MNRAS.418..838W} results in reasonable estimates of column densities. \citet{1977ApJ...216..713K} have found that molecular hydrogen will be dissociated by shocks with shock-speeds exceeding $24\,\text{km}\,\text{s}^{-1}$. The typical shock speeds in superbubbles exceed $100\,\text{km}\,\text{s}^{-1}\,$ and thus the molecular hydrogen in the post-shock region would almost certainly be dissociated. In our feedback model we are not treating the chemical evolution associated to SN shocks and thus we might overestimate the molecular abundance in the shocked gas. We address this possible inconsistency by neglecting the contributions of particles hotter than $10^{4}\,\text{K}\,$ to the molecular hydrogen budget.

\subsection{The Interstellar Radiation Field}
\label{sec:ISRF}

In order to compute the photo-dissociation, photo-heating and photo-ionization rates due to the ISRF, we take the spectral energy distribution of a simple stellar population from 
Starburst99\footnote{\label{footer:Starburst99} \href{https://www.stsci.edu/science/starburst99/docs/default.htm}{https://www.stsci.edu/science/starburst99/docs/default.htm}} \citep{1999ApJS..123....3L}. 
We use the time averaged spectrum for various stellar metallicities within the first $4 \,\text{Myr}\,$ of the stellar population's lifetime, calculated using spectra at equidistantly spaced times with a timestep of $\Delta t = 1\,\text{Myr}\,$.

We use these spectra to compute the ratio of the reaction rates for the photo-ionization of H, He and H$_2$ to the H$_2$ photo-dissociation rate.
The reaction rates are computed using 
\begin{equation}
    \Gamma = \frac{1}{4\pi r^2} \int_{\nu_\mathrm{th}}^{\infty} L_{\nu} \sigma_\nu \frac{d\nu}{h \nu}.
\end{equation}
We compute the photo-heating rate as 
\begin{equation}
    \mathcal{H} = \left\langle\epsilon\right\rangle\Gamma,
\end{equation}
where the average emitted energy per reaction is given by
\begin{equation}
    \left\langle\epsilon\right\rangle = \frac{1}{4\pi r^2 \Gamma} \int_{\nu_\mathrm{th}}^{\infty} L_{\nu} \sigma_\nu \left(h \nu - h \nu_\mathrm{th}\right) \frac{d\nu}{h \nu}.
\end{equation}
Here $h$ is the Planck constant, and $h \nu_\mathrm{th}$ is the ionization threshold energy. The photo-ionization cross-sections $\sigma_{\nu}$ are taken from Table~3 of \citet{1987ApJ...318...32S}, and the photo-dissociation cross-section is taken from \citet{1997NewA....2..181A}. 
The spectrum is independent of metallicity in the LW-band ($\lambda \sim 10^3\, \text{\AA}\,$), while it exhibits some metallicity dependency at shorter wavelengths.
As a result, the $\text{H}_2$ photo-dissociation rate and the typical heating energies are essentially independent of metallicity $Z$, while the ratios of the rates exhibit a mild power law dependence:
\begin{equation}
    \frac{\Gamma}{\Gamma_{\text{H}_2}} \sim \frac{\Gamma_{\odot}}{\Gamma_{\text{H}_2}} \left(\frac{Z}{Z_{\odot}}\right)^{\alpha}.
\end{equation}

\begin{table}
\caption{\label{table:photoionisation} Photo-ionisation parameters}
\begin{tabular}{c|c c c}
Reaction & $\left\langle\epsilon\right\rangle\left[\mathrm{eV}\right]$ & $\Gamma/ \Gamma_{\text{H}_2}$ & $\alpha$\\
\hline
$\text{H} + \gamma \rightarrow \text{H}^+ + \text{e}^-$ & $3.673$ & $0.518$ & $-0.100$\\
$\text{He} + \gamma \rightarrow \text{He}^+ + \text{e}^-$ & $5.389$ & $0.156$ & $-0.286$\\
$\text{H}_2 + \gamma \rightarrow \text{H}_2^+ + \text{e}^-$ & $4.566$ & $0.805$ & $-0.135$\\
\end{tabular}
\end{table}

The results of the calculation are displayed in Table~\ref{table:photoionisation}. Now, in order to compute the strength of the photo-ionization, heating and dissociation, the strength of the ISRF needs to be estimated. We have devised two separate methods for this. The first method which is based on the treatment of early stellar feedback (ESFB) employed in the {\sc Gadget3-Osaka} feedback model \citep{2019MNRAS.484.2632S} assumes that the radiation is injected into the gas surrounding a star particle at a fixed number of equidistantly spaced times before the onset of the first type II SN events, causing sudden photo-ionization, heating and dissociation. The other method based on the prescription of \citetalias{2018MNRAS.474.1545C} continuously accounts for the effect the ISRF has on the gas, by tracing the ISRF using the star formation column density $\Sigma_\text{SFR}$. In the following the two methods are described in more detail.

\subsubsection{Early Stellar Feedback}\label{sec:ESFB}
In the model for ESFB described by \citet{2019MNRAS.484.2632S}, the effect of the ISRF is modelled by heating up the surrounding gas at $N_\text{ESFB} = 8$ equidistant times before the onset of the first type II SN event. In their model the energy is distributed to the surrounding gas particles in SPH fashion as a kernel-weighted sum. A problem with this treatment is that if $N_\text{ESFB}$ is too large and too few neighbors are considered, an individual particle can be injected with unphysically large amounts of energy, which could never be achieved through photo-ionization of the atomic gas present in the gas particles. 
In order to remedy this, \citet{2019MNRAS.484.2632S} limit the maximum achievable temperature due to ESFB to $T = 2 \times 10^4\, \text{K}\,$. This limiting procedure however is unphysical and can lead to the loss of large amounts of energy, that might otherwise be injected elsewhere. 

A different approach is to consider the atomic gas present in the gas particles as `fuel' instead and treat the heating as a consequence of its ionization. This way the maximum achievable temperatures are naturally limited to few~$\times 10^4\, \text{K}\,$  by the amount of available `fuel' that can be ionized. To this end, we estimate the amount of ionizations of H atoms by assuming that all of the injected energy goes into photo-ionization of {\sc H i}. The resulting energy balance for a gas particle $j$ with {\sc H i} mass fraction $X_{\text{HI}, j}$ reads
\begin{equation}\label{eq:EnergyBalance}
    X_{\text{HI}, j}\, \omega_j \Delta E_{\text{ESFB}} = \left(\epsilon^{\rm ion}_{\text{HI}} + \left\langle\epsilon_{\text{HI}}\right\rangle \right) N^{\rm ion}_{\text{HI}, j},
\end{equation}
where $X_{\text{HI}, j} \omega_j \Delta E_{\text{ESFB}}$ is the fraction of the energy output of the feedback event assigned to the gas particle $j$, $\epsilon^{\rm ion}_{\text{HI}} = 13.6 \,\text{eV}\,$ is the energy required to ionize a hydrogen atom, and $N^{\rm ion}_{\text{HI}, j} = \frac{m_j}{m_\text{H}} dX_{\text{HI}, j}$ is the number of ionizing reactions occuring in gas particle $j$ due to the incident radiation. We parametrize the weight $X_{\text{HI}, j} \omega_j$ on the left-hand-side of equation (\ref{eq:EnergyBalance}) to be proportional to $X_{\text{HI}, j}$, noting that the amount of energy that can be absorbed by a gas particle is limited by the available amount of neutral gas `fuel', which can be ionized by the incident radiation.
The resulting amount of ionization is
\begin{equation}
    \frac{dX_{\text{HI}, j}}{X_{\text{HI}, j}} \sim -750 \left(\frac{E_\text{bol}}{2.0 \times 10^{50}\,\text{erg} / \text{M}_{\sun}\,}\right) \left(\frac{8}{N_\text{ESFB}}\right)\omega_j \frac{m_*}{m_j},
\end{equation}
where $E_\text{bol} = 2.0 \times 10^{50}\,\text{erg} / \text{M}_{\sun}\,$ is the specific bolometric energy output in UV radiation ($\epsilon > 13.6\,\text{eV}\,$) and $m_{*}$ is the mass of the star particle. Given that $\frac{dX_i}{X_i} = \Gamma_i dt$, we derive the ionization/dissociation of the other elements using the previously established ratios of the rates. The thus obtained ionization rates are appropriately attenuated due to dust and self-shielding. Once the amount of ionization of each species has been computed, the heating energy is injected by adding up the contributions from each species. It is apparent that in order for this method to make sense, $\left|dX/X\right| < 1$ has to be satisfied.
This requires extremely large numbers of neighboring gas particles included in the feedback calculation and large numbers of feedback events per star, leading to prohibitively short timesteps.
Due to these limitations, we consider this method to be of very limited use in practice, but for illustration still include a run where we apply this model with 128 neighbors and $N_\text{ESFB} = 8$, noting that we expect rather large amounts of energy lost due to gas particles with $\left|dX/X\right| \gg 1$.

\subsubsection{ISRF from SFR}\label{sec:ISRFfromSFR}

In the above model for the radiation feedback, one of the main limitations is that unless feedback events are occuring frequently enough, rapid recombination occuring on similarly short timescales can immediately undo the ionization and heating due to the ESFB. Therefore it would be desirable if the strength of the ISRF could be estimated from local properties of the gas, allowing to include its effects on heating and chemistry in the non-equilibrium chemistry network and integrating them consistently with all other reactions. To this end, \citet{2005MNRAS.356.1529H} derived a proportionality of the strength of the ISRF $\chi$ in the LW-band with the star formation rate column density as 
\begin{equation}
    \chi = \frac{\Sigma_{\text{SFR}}}{1.7\times 10^{-3}\, \text{M}_{\sun}\,\text{yr}^{-1}\,\text{kpc}^{-2}\,}.
\end{equation}
Here $\chi$ is measured in units of the solar neighborhood value derived by \citet{1968BAN....19..421H} $3.2 \times 10^{-20}\,\text{erg}\,\text{s}^{-1}\,\text{cm}^{-2}\,\text{Hz}^{-1}\,\text{sr}^{-1}$, $\Sigma_{\text{SFR}}$ is determined by computing an SPH estimate of the star formation rate density $\Dot{\rho}_{\text{SFR}}$ and multiplying it with the local Sobolev-like length scale $L_{\text{sob}}$.
This model further enables us to include photo-electric heating and recombination cooling on grain surfaces in {\sc Grackle}, as these processes require the user to provide the strength of the ISRF. In the \ESFB\ run, we neglect these processes, since we do not estimate the strength of the ISRF. It should be noted that we are not accounting for the grain size dependence of these processes, which are calculated by {\sc Grackle}. As these processes occur on grain surfaces, neglecting the grain size dependence might potentially affect our results. Here we argue why the effect this might have is likely small. There are three main ingredients to these processes: the strength of the ISRF, the total dust abundance and the shape of the GSD. If the ISRF is weak both processes can be neglected,
we are only concerned with the epoch with strong star formation.
In our setup, there is some strong star formation at early times, but during this stage there is only little dust and therefore both rates should be rather small compared to other heating and cooling rates. At later times there is more dust, but the star formation activity is declining and so is the ISRF. Furthermore since the GSD approaches the MRN distribution in the dense star-forming ISM at later times, the error we make by neglecting the shape of the GSD becomes smaller over time.

\subsection{Simulation Suite}\label{sec:runs}

\begin{table}
\caption{List of different simulations}
\label{tab:runs}
\begin{tabular}{lc}
\hline
Run name & Comment\\
\hline
Vanilla & ISRF from SFR\\
NoISRF & No ESFB or ISRF\\
ESFB & ESFB instead of ISRF from SFR\\
NoGSD & Grain-size-independent $R_{\text{H}_{2}}$ and $\tau_{\text{LW}}$\\
NoSh & Self-shielding turned off\\
NoPr & H$_2$ formation on grains turned off\\
\hline
\end{tabular}
\end{table}

In order to emphasize the importance of each process in different circumstances, we have run different simulations where details in the used models for the processes related to the formation and dissociation of molecular hydrogen and the ISRF have been modified. Dust evolution, gas phase chemistry and the UVB are kept the same in all runs. A list of all runs and their characteristics is given Table~\ref{tab:runs}. In the following we will give a more detailed description of each run.

We refer to our fiducial model as the \Vanilla\ model. In this run we consider formation of H$_2$ on grain surfaces and dissociation of molecules due to the ISRF, which we model based on the prescription described in section \ref{sec:ISRFfromSFR}. Self-shielding and dust extinction are employed as described in section \ref{sec:dissociation}.
In order to provide a baseline for the impact of the ISRF, in the \NoISRF\ run we turn off all processes associated to the ISRF. All other processes are the same as in the \Vanilla\ model. The only difference between the \ESFB\ run and the \Vanilla\ run is, that we replace the continuous prescription for the ISRF based on the local SFR column density used in the \Vanilla\ model with an event-based one as described in section \ref{sec:ESFB}. In order to quantify the effect grain size has on the molecular abundance, in the \NoGSD\ model, we replace the grain-size-dependent expressions in eqs. \ref{eq:formationGSD} and \ref{eq:extinctionGSD} with the grain-size-independent ones in eqs. \ref{eq:formationNoGSD} and \ref{eq:extinctionNoGSD}. In the \NoSh\ run we disable self-shielding in order to appreciate the importance of self-shielding. Finally, in \NoPr\ we disable H$_2$ formation on grain surfaces to estimate the impact this process has on molecule formation.

\section{Results}\label{sec:results}

We study the importance of different processes for the formation of molecular hydrogen in simulations of an isolated Milky-Way-like galaxy. In order to emphasize the importance of each process in different circumstances, we have run different simulations where individual processes have been turned off or details in the implementation have been modified compared to the fiducial case, as described in section \ref{sec:runs}. A list of all runs and their characteristics is given in Table~\ref{tab:runs}.

We first study the global formation history of the molecular gas component and its spatial distribution. We then take a look at a number of scaling relations between the molecular gas fraction and a number of related observables.

\subsection{Star Formation and Enrichment History}

\begin{figure}
\includegraphics[width=0.45\textwidth]{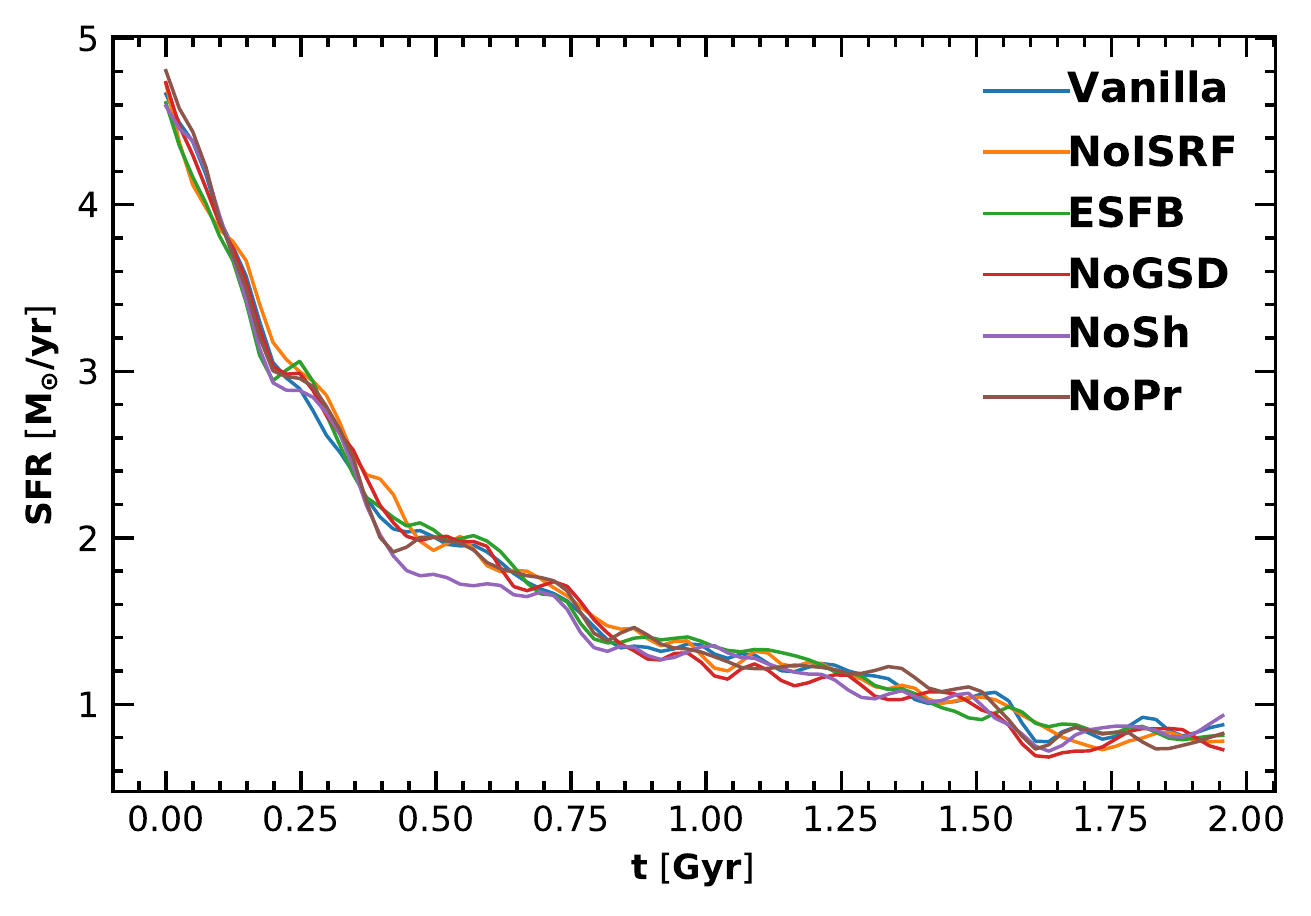}
\caption{The starformation history of the simulated galaxy over the simulated timespan for the different models. All runs have a similar star formation history. Different runs are indicated by different colours as shown in the legend, and the same set of colours is used for other figures.}\label{fig:SFhistory}
\end{figure}
\begin{figure}
\includegraphics[width=0.45\textwidth]{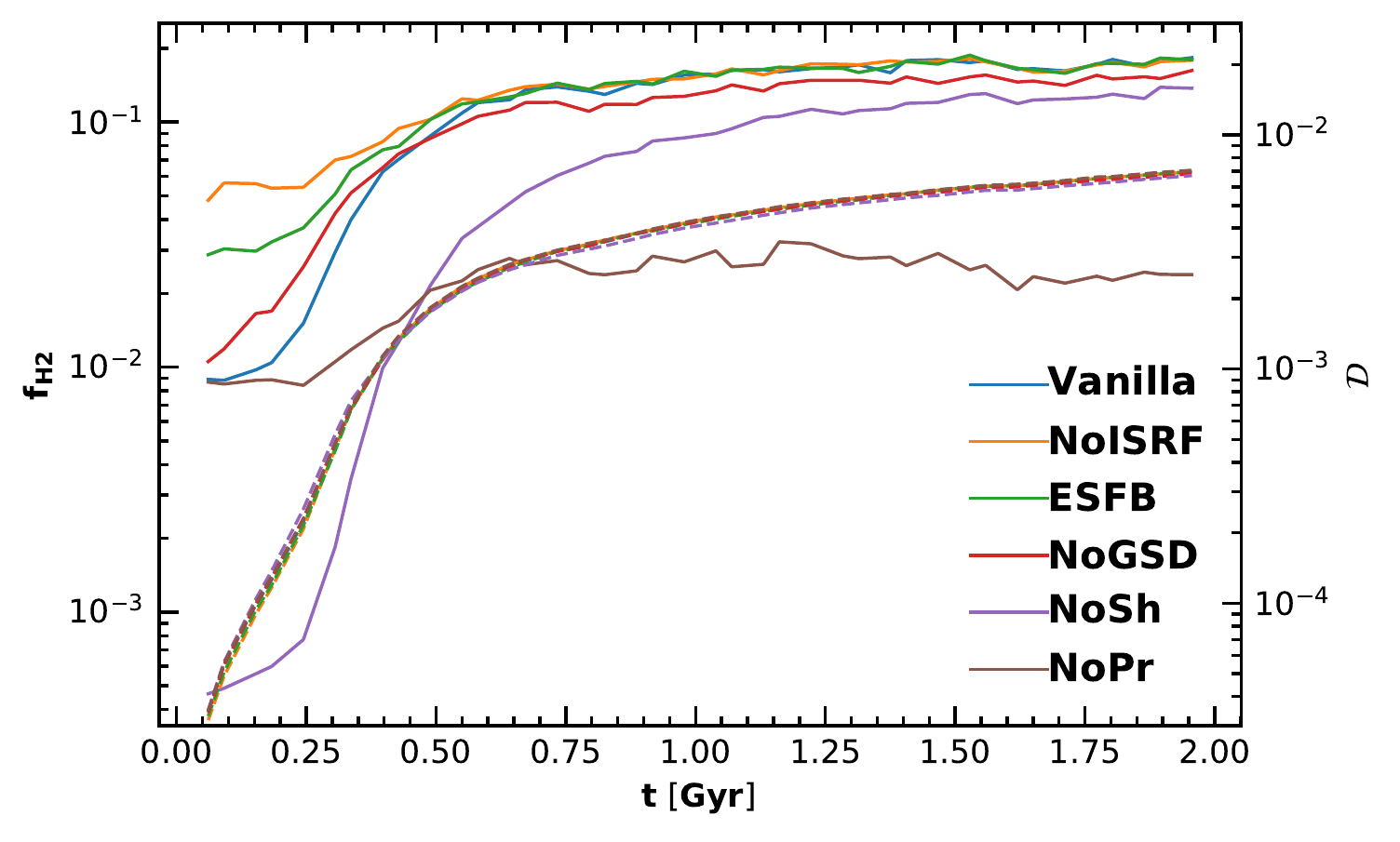}
\caption{The galactic molecular fraction (solid lines; left axis) and dust-to-gas ratio (dashed lines; right axis) as a function of time for the different models.}\label{fig:H2history}
\end{figure}
The processes linked to the formation and destruction of H$_2$ depend on the strength of the ISRF and the abundance of dust grains, both of which are connected to the star formation history of the galaxy. Thus, in order to ensure a fair comparison between the runs, it is important to verify that the different models do not modify the star formation history too much. In Figure~\ref{fig:SFhistory} we show the galactic star formation rate as a function of time for the different runs. As one can see, the starformation history is roughly the same in all runs. We can thus move on and compare their results.

In Figure~\ref{fig:H2history} we show the galactic molecular gas fraction and dust-to-gas ratio as a function of time for each model. In all runs except in \NoPr\ run, the molecular fraction is low at first, before it starts to grow as the galaxy becomes enriched with dust and then saturates once the dust growth slows down. At early times, the molecular abundance is dictated by a competition of formation in the gas phase and dissociation due to the initially strong ISRF. The differences between the models arise here due to differences in the treatment of the ISRF or the treatment of self-shielding, which can reduce the impact of the ISRF. 
Accordingly, the molecular abundance is the highest in the run without ISRF, where the only source of dissociation is the UVB. Ideally the runs with source-based (\ESFB) and sink-based (\Vanilla) ISRF should lead to identical results, but the molecular abundance in the \Vanilla\ run is lower by a factor of $\sim 3$, indicating that the feedback due to the ISRF is weaker in the \ESFB\ run. In the run where we ignore the grain size dependence of dust shielding and formation of H$_2$ on dust grains, the molecular fraction is slightly higher than in the Vanilla run, which is likely because the GSD is initially dominated by large grains which have a smaller surface area and thus processes like extinction or formation on grain surfaces are yet inefficient. Since the grain-size-independent model implicitly assumes a fixed GSD, the amount of small grains and thus their effect is initially overestimated. In the run without self-shielding, the ISRF effectively dissociates molecules leaving behind a tiny molecular fraction.

Only much later as the star formation rate and with it the strength of the ISRF decreases and the disk becomes enriched with dust grains which can shield the molecules from the residual radiation and act as catalyst for the synthesis of H$_2$, significant molecular abundances can be reached. 
In the runs with formation on dust and self-shielding, the time at which the molecular fraction begins to saturate is around $t \sim 500 \,\text{Myr}\,$, while this is slightly delayed in the run without self-shielding. In the run without formation on dust grains, the molecular fraction at saturation is only at around $2\,\%\,$, while it is up to an order of magnitude larger if it is included. Without self-shielding, the molecular fraction is slightly lower than in the other runs, owing to the stronger dissociation. In the run, where the grain size dependence is ignored, the final value is also slightly reduced, possibly due to slight differences in the normalisation of the formation rate and the optical depth.

\begin{figure*}
\includegraphics[width=0.95\textwidth]{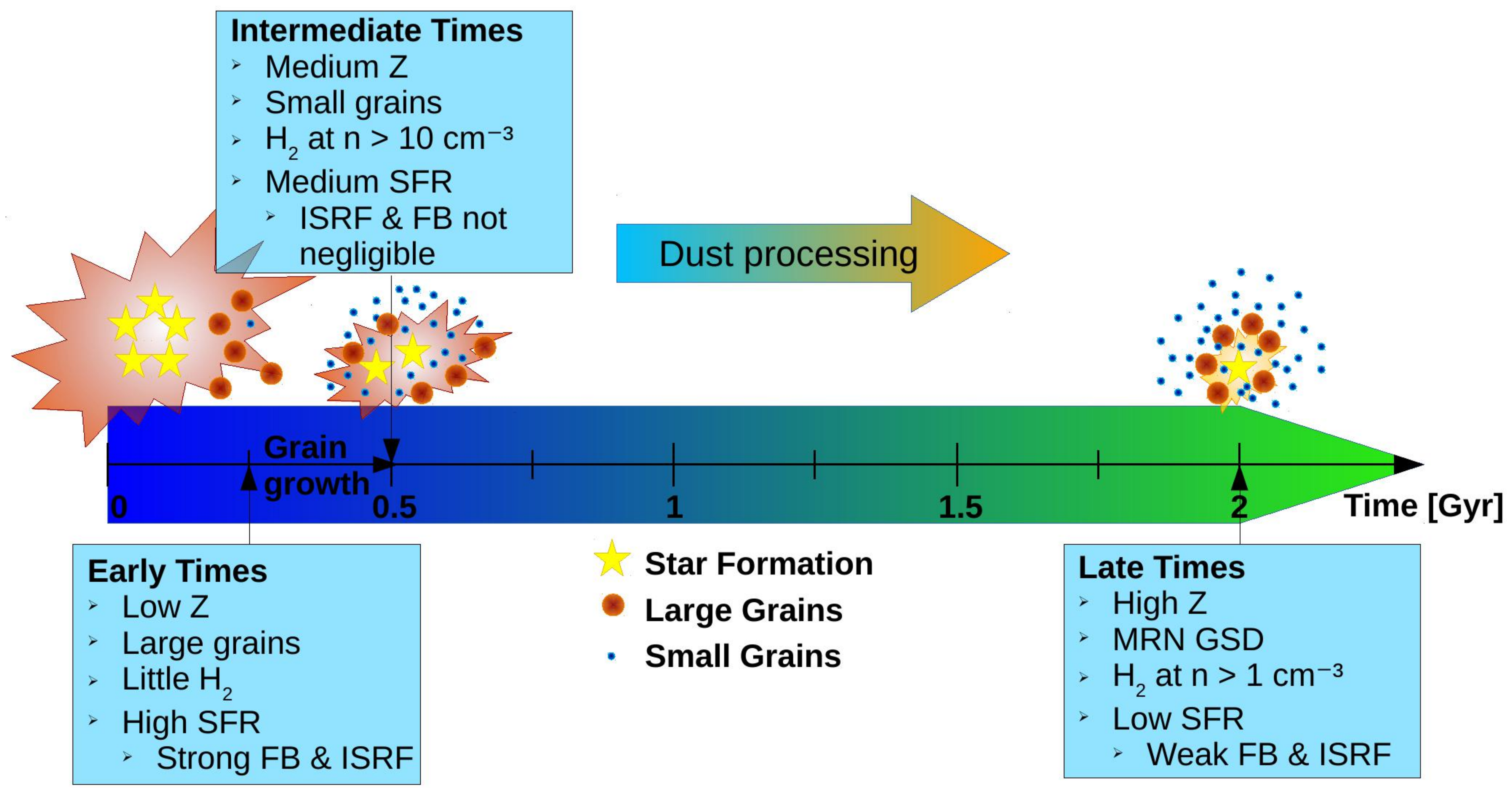}
\caption{Schematic overview of the epochs of dust and molecular evolution in our simulations. The number of stars is indicative of the star formation rate. The number of brown and blue dots is indicative of the abundance of large and small dust grains, respectively. The "explosion" feature surrounding the stars indicates stellar feedback. The colorbar on the time axis corresponds qualitatively to the color of the galaxy, i.e. blue corresponding to many young stars and green corresponding to older stars.}
\label{fig:timeline}
\end{figure*}

This evolution of molecular hydrogen and dust depicted in the figure indicate that there are three interesting epochs in the evolution of the galaxy, where the distribution of molecular hydrogen might have very different properties. The first epoch corresponds to the first few 100 Myr, where the abundance of molecular hydrogen is dominated by the strong ISRF. The next epoch is located at the position of the turnover from exponential dust growth to the more slow enrichment at around $t \sim 500 \,\text{Myr}\,$. At this point molecular enrichment slows down but the molecular fraction has not yet reached its final value. Finally, the last epoch is located at the end of our simulated period, where the molecular abundance is roughly constant and the GSD has settled to a distribution resembling the \citetalias{1977ApJ...217..425M} GSD. A schematic overview of the timeline is given in Figure~\ref{fig:timeline}. In order to study the properties of the molecular gas in these epochs, we show all following figures at $t = 200\,\text{Myr}\,$, $t = 500\,\text{Myr}\,$ and $t = 2\,\text{Gyr}\,$ and refer to them as early, intermediate and late times, respectively.

\subsection{Spatial Distribution of Molecular Gas}

\begin{figure*}
\includegraphics[width=0.95\textwidth]{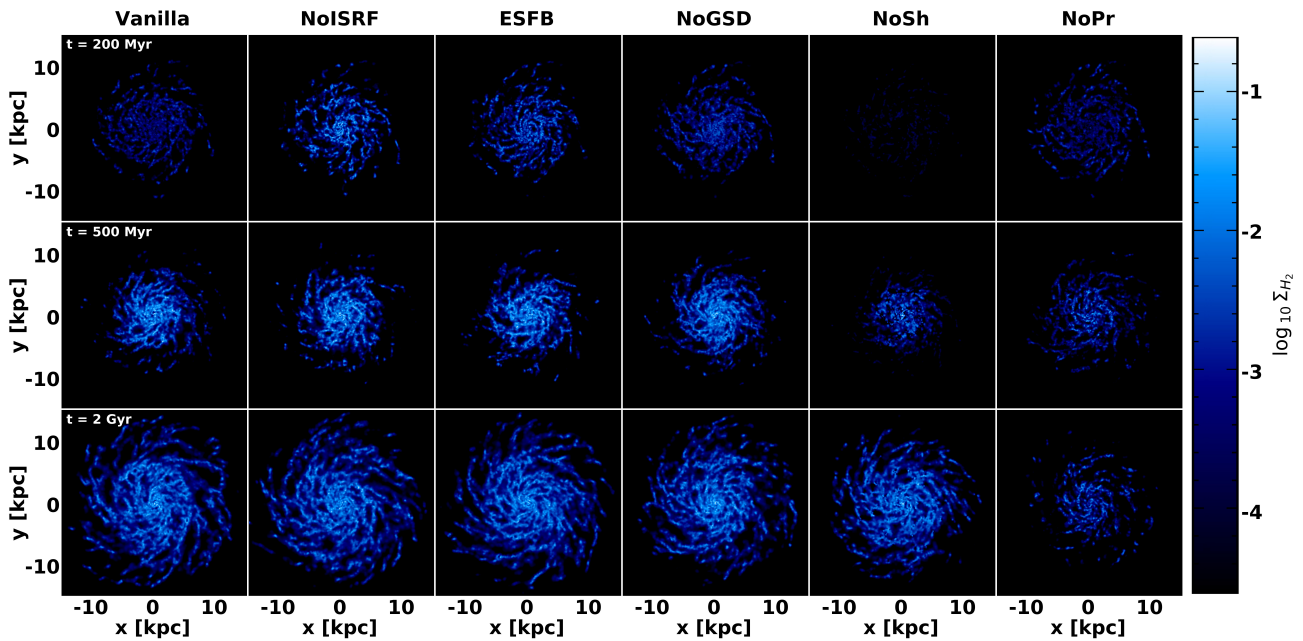}
\caption{Face-on projection map of molecular hydrogen column density for the different runs at different snapshots in a $30\,\text{kpc}\,$ box. Top panels show the H$_2$ column density at early times ($t = 200\,\text{Myr}\,$), middle panels at intermediate times ($t = 500\,\text{Myr}\,$) and bottom panels at late times ($t = 2\,\text{Gyr}\,$.) Different runs are shown in different columns with the name of the runs shown on the top.}
\label{fig:H2map}
\end{figure*}

The processes that determine the abundance of molecular hydrogen depend on the local physical conditions. Thus excluding them or modifiying their implementation may have an impact on the spatial distribution of molecular hydrogen. To explore this point, we show spatial maps of the molecular hydrogen column density in Figure~\ref{fig:H2map}. The projections have been made with the SPH visualization code SPLASH \citep{2007PASA...24..159P}.

In all runs except \NoPr\ and \NoSh, the molecular gas is initially confined to the inner $\sim 10\,\text{kpc}\,$ of the disk and roughly stays within these bounds at intermediate times. At late times it extends much further into the outer parts of the disk up to almost $15\,\text{kpc}\,$. In \NoPr\ the abundance hardly evolves from early to late times and stays roughly confined within the inner $10\,\text{kpc}\,$. In \NoSh\ the evolution is delayed. At early times, the molecular column density falls below the lower bound of the color scale almost everywhere. At intermediate times it begins to grow in the center and seems to have almost caught up with the distribution in the \Vanilla\ run at late times. In the other runs, the spatial distribution is similar, but at early times, the abundances are slightly lower in the \Vanilla\ and \NoGSD\ runs as compared to the \NoISRF\ and \ESFB\ runs.

\begin{figure}
\includegraphics[width=0.45\textwidth]{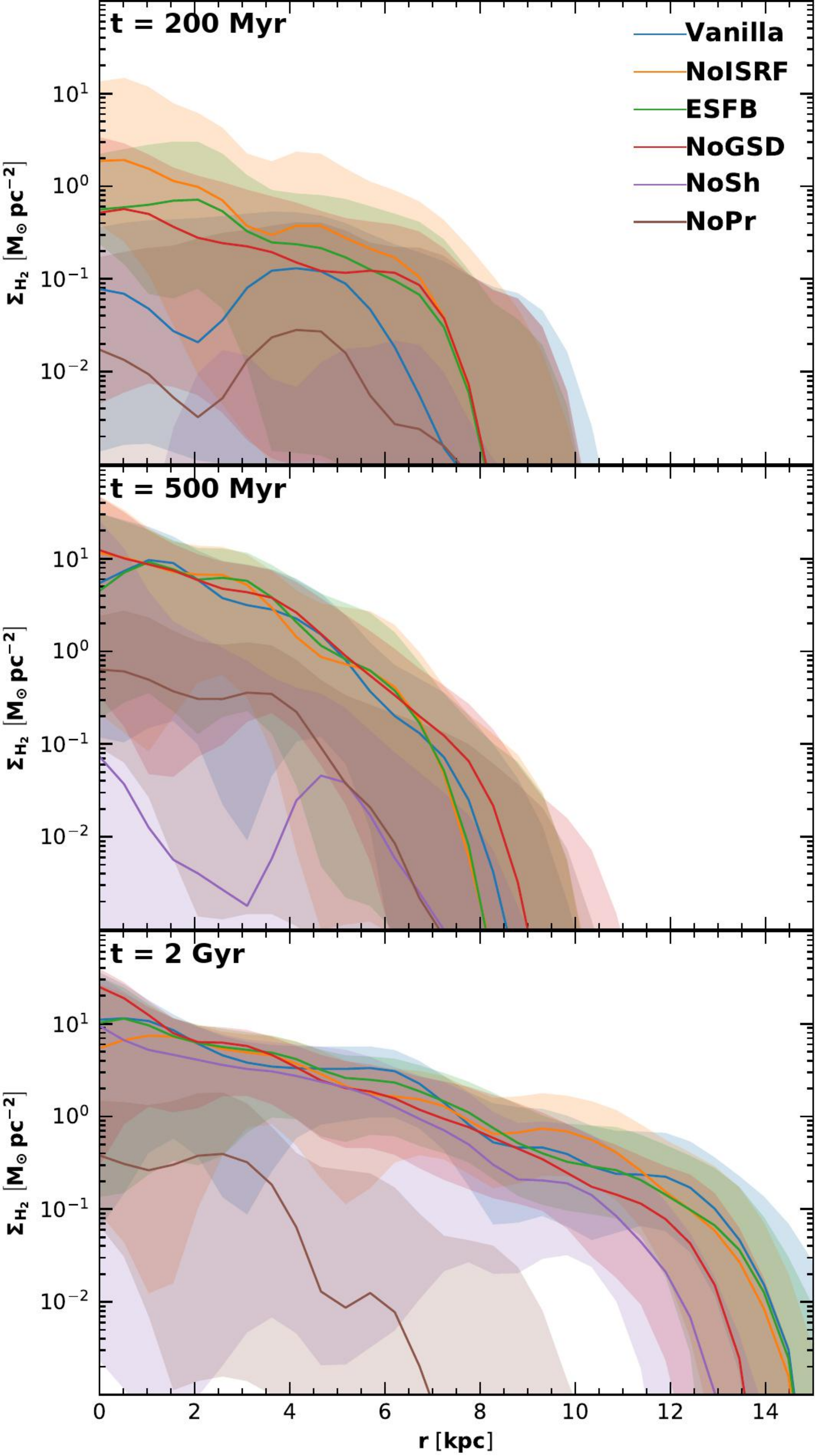}
\caption{The molecular column density as a function of cylindrical radius $r$ for the different simulations at early (top) intermediate (center) and late times (bottom). Solid lines depict the median relation, while the shaded area shows the range of values between the 25th and 75th percentile.}\label{fig:RadialH2}
\end{figure}

In order to further quantify the differences between the models, seen in Figure~\ref{fig:H2map}, we show the column density of H$_2$ as a function of cylindrical radius $r$ in Figure~\ref{fig:RadialH2}. We compute the column density by projecting particles onto circular annuli of equal width $\text{d}r = 500 \,\text{pc}\,$ between $r = 0$ and $r = 15\,\text{kpc}\,$, where we define $r$ with respect to the density weighted center-of-mass of the gas particles.

At early times, molecular hydrogen is confined within $r < 7.5\,\text{kpc}\,$ in all runs. The column density profiles exhibit a shallow decline from the central value of $\Sigma_{\text{H}_2} \sim 1 \,\text{M}_{\sun}\,\text{pc}^{-2}$ in the runs \NoISRF, \ESFB\ and \NoGSD\ and then rapidly drop beyond $r \sim 7\,\text{kpc}\,$. In \Vanilla\ and \NoPr\ the profile is rather flat with $\Sigma_{\text{H}_2} \sim 10^{-2} - 10^{-1}\, \text{M}_{\sun}\,\text{pc}^{-2}$ and drops more slowly towards large radii. In \NoSh\ the column density is lower than the axis limits. In all runs there is a lot of scatter, due to the fact that feedback events and star formation can locally deplete all molecules. 

At intermediate times, the molecular column density profiles have steepened and there are fewer differences between the runs that include self-shielding and formation of molecular hydrogen on dust grains. In these runs, levels of H$_2$ at $r \sim 7.5\,\text{kpc}\,$ are still similar to early times, and the drop beyond this radius is still equally steep, but the central value of the column density has now increased by an order of magnitude compared to early times. In the run without formation on grains, the profile is more centered towards smaller radii and reaches column densities that are about an order of magnitude lower than in the runs that include it (and self-shielding). In the run without self-shielding, the central value of the column density is comparable to the other runs with formation of H$_2$ on grains, but it rapidly drops and reaches column densities that are two orders of magnitude below the ones in the \NoPr\ run at intermediate radii, before it increases again at $r \sim 5\,\text{kpc}\,$, where it becomes comparable to the column density in the \NoPr\ run. It should be noted however, that there is significant scatter in the profile of the \NoSh\ run, indicating that this trend might just be a statistical effect. 

At late times, the profile in all runs except for the \NoPr\ run extends to larger radii. Molecular column densities of $\Sigma_{\text{H}_2} \sim 0.1\, \text{M}_{\sun}\,\text{pc}^{-2}$ are reached up to radii of $r \sim 12 - 14\,\text{kpc}\,$, while the central column densities are still comparable to the ones at intermediate times. In \NoSh\ and \NoGSD, the rapid drop of the column density occurs at slightly smaller radii. In \NoSh\ this is likely because H$_2$ formation has been slower and still needs to catch up. In \NoGSD\ the difference might arise, because at large radii where gas is warmer and more diffuse, shattering is more efficient, leading to a peak in the small-to-large-grain ratio. This would allow for more efficient net H$_2$ production in the runs with grain-size-dependent extinction and formation on grain surfaces, making higher H$_2$ column densities possible at larger radii. In the run without formation of H$_2$ on grains, the profile is similar to the one at intermediate times. This is because it essentially obeys equilibrium with formation and destruction processes in the gas phase, which are inefficient at the densities resolved by the simulation.

\subsection{Scaling Relations}

\subsubsection{Hydrogen density}\label{sec:H2vsn}

\begin{figure}
\includegraphics[width=0.435\textwidth]{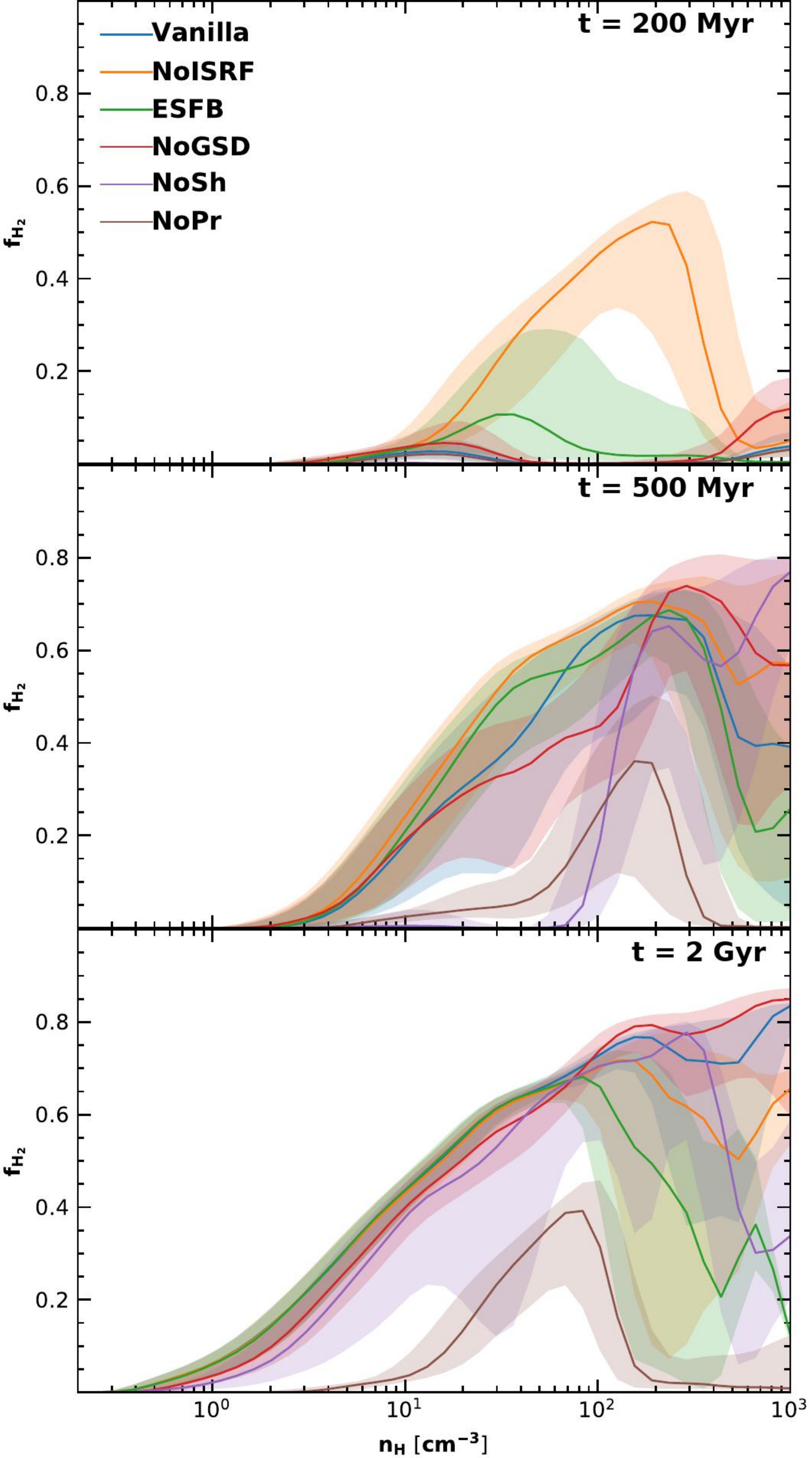}
\caption{The molecular fraction as a function of hydrogen density at early (top), intermediate (center) and late (bottom) times for the different runs. Solid lines depict the median relation, while the shaded area shows the range of values between the 25th and 75th percentile.}\label{fig:H2vsn}
\end{figure}

\begin{figure}
\includegraphics[width=0.435\textwidth]{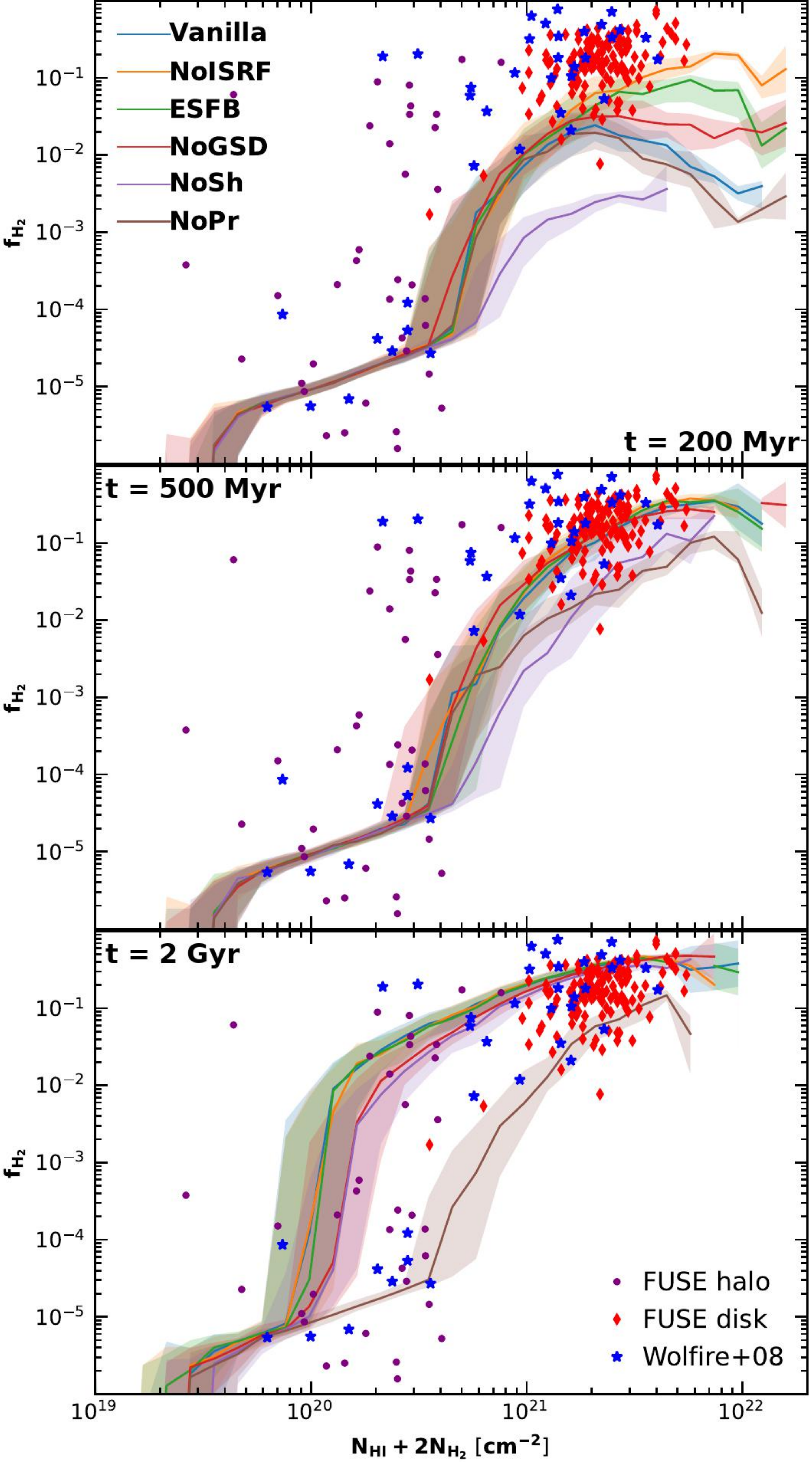}
\caption{Same as Figure~\ref{fig:H2vsn}, but for the molecular fraction as a function of hydrogen column density. Data points correspond to measurements in the Milky Way disk and halo taken from \citet{2008ApJ...680..384W} and the FUSE survey \citep{2006ApJ...636..891G, 2021ApJ...911...55S}.}\label{fig:H2vsNH}
\end{figure}

In Figure~\ref{fig:H2vsn}, we show the relation between the molecular fraction and the SPH hydrogen density. It is important to note, that we assume that a fraction of the gas is condensed into dense clouds on subgrid scales as described in section \ref{sec:h2production}.

At early times, the molecular fractions are low (\fmol$ \lesssim 10\,\%$) at all densities, in all runs except in the run without ISRF. In this run, the molecular fraction increases at densities above $10\,\text{cm}^{-3}\,$, reaches a maximum at $n_\text{H} \sim 200\,{\rm cm}^{-3}\,$, and then drops at larger densities. The shape of the relation is determined by formation and dissociation of H$_2$ in the gas phase. At high densities the molecular fraction drops due to the influence of stellar feedback, as high density gas tend to be located in the vicinity of newly formed stars .

At intermediate times, in all models density regimes with significant H$_2$ production exist. In the runs with self-shielding and formation on grain surfaces, the molecular fraction starts to rise above the percent level at densities $n_\text{H} \gtrsim 2\,\text{cm}^{-3}\,$ and saturates at $n_\text{H} \gtrsim 100\,\text{cm}^{-3}\,$ with molecular fractions of the order \fmol$ \sim 60\,\%\,$, even though in all cases the median molecular fraction starts to drop at this point, with a large scatter.
In \NoGSD\ the molecular fraction is lower than in the other runs at densities $n_{\text{H}} \sim 10\,\text{cm}^{-3}\,$, probably because of enrichment with small grains due to efficient accretion of gas phase metals. \ESFB\ tends to be most efficient in the densest regions, lowering the molecular fraction at densities
$n_{\text{H}} \sim 600\,\text{cm}^{-3}\,$. In the run without ISRF, the molecular fractions tend to be slightly higher than in the runs including it. In the run without self-shielding, the molecular fraction is low up to densities of $n_{\text{H}} \sim 70\,\text{cm}^{-3}\,$ above which it rapidly increases, due to the presence of dust, which becomes the dominant source of shielding. In \NoPr\ the molecular fraction follows a similar trend to the relation at early times in the \NoISRF\ run, though with a slightly lower peak value, resulting from photodissociation due to the ISRF.

At late times all runs except \NoPr\ have molecular fractions exceeding the percent level above hydrogen densities $n_{\text{H}} \gtrsim 0.3\,\text{cm}^{-3}\,$. In the \Vanilla, \NoISRF\ and \ESFB\ runs, the relation starts to rise at slightly lower densities than in the \NoGSD\ and \NoSh\ runs and the molecular fractions tend to be higher up to $n_{\text{H}} \sim 50\,\text{cm}^{-3}\,$, above which the relation either drops (\ESFB, \NoISRF, \NoSh) or saturates (\Vanilla, \NoGSD). The reasons for the drop in the \NoSh\ and \ESFB\ runs are essentially that dissociation due to the ISRF is strongest in dense gas in these models. The relation in the \NoPr\ run is similar to that at intermediate times, but the bump is extending towards lower densities, as star formation is slowing down and extinction due to dust grains becomes a more dominant source of shielding at lower densities.

In Figure~\ref{fig:H2vsNH}, we compare the molecular fraction as a function of hydrogen column density to observational data in the disk and halo of the Milky Way galaxy. We compute the column densities from our simulations, by dividing the galactic plane into a grid with pixels of side length $500\,\text{pc}\,$ and projecting the desired quantities of the particles within a scale height of $\left|z\right| < 3\,\text{kpc}\,$ onto the grid. We then exclude all grid cells with vanishing $\Sigma_{\text{HI + H}_2}$ and compute the molecular fraction in each remaining cell as the ratio of the molecular column density to the total hydrogen column density. We assign the cells to 30 logarithmically spaced bins on the $N_{\text{HI}} + 2N_{\text{H}_2}$ axis to compute the median and 25th (75th) percentile in each bin. The blue stars correspond to the observations in the disk of the Milky Way from \citet{2008ApJ...680..384W}, the purple dots to observations of the Milky Way Halo from the FUSE survey \citep{2006ApJ...636..891G} and the red diamonds to observations of the Milky Way disk from the FUSE survey \citep{2021ApJ...911...55S}.

The molecular fraction generally correlates positively with the hydrogen column density. 
At low column densities it slowly increases with typical values of \fmol$\sim 10^{-5}\,$ until a threshold column density is reached above which it jumps by three orders of magnitude, taking values exceeding $1\,\%$ and even reaching values exceeding $10\%$ at the highest resolved column densities. The value of the threshold column density is lowered at late times in all runs, except \NoPr\ where it stays at the initial value of $N_{\text{HI}} + 2 N_{\text{H}_{2}} \sim 2\times 10^{20}\,\text{cm}^{-2}\,$. The lowering of this value can thus be understood as a result of molecular formation on grain surfaces, which enhances the molecular fraction even at low surface densities as the diffuse gas gets enriched with dust.

At early times there are some differences between the models at high column densities. The steep rise in molecular fractions is less pronounced in the run without self-shielding, reaching molecular frations of only about \fmol$\sim 10^{-3}\,$. In the other runs, each run exhibits a different trend towards very high column densities. \Vanilla\ and \NoPr\ exhibit a slight drop, in \NoGSD\ the molecular fraction is roughly constant and in \ESFB\ and \NoISRF\ it increases towards the highest resolved column densities. These differences arise due to differences in the treatment of the ISRF and shielding from it. In the run without the ISRF, the absence of the ISRF allows for high molecular fractions even in the presence of many young stars, which provides a base line for the different treatments of radiation feedback. In the \ESFB\ model the molecular fraction is lowered much less than in the \Vanilla\ model, indicating that the event based radiation feedback is much less efficient in keeping molecular abundances low. Only at very high column densities there is a slight dip in the molecular fraction, indicating that ESFB is only active on very short distances from the stars, but does not reach out much further. The molecular fractions in \Vanilla\ are only slightly higher than in \NoPr, indicating that molecule formation on grain surfaces is a subdominant process at early times. In the \NoGSD\ run, the molecular fraction is much higher than in the \Vanilla\ run at the highest column densities. This is because \NoGSD\ implicitly assumes an \citetalias{1977ApJ...217..425M} GSD,
while the actual GSD at early times is dominated by large grains which hardly contribute to grain surface processes, leading to an overestimate of molecule formation and dust shielding. 

At intermediate times most runs have settled to an almost identical relation between the molecular fraction and the hydrogen column density. Only \NoSh\ and \NoPr\ result in lower molecular hydrogen abundances at high column densities. 

At late times the enrichment of diffuse gas with dust makes it possible for high molecular fractions to be maintained at low column densities. The formation of molecular hydrogen is less efficient in \NoSh\ and \NoGSD.  In the \NoSh\ run, the absence of self-shielding leads to more photo-dissociation and thus lower molecular fractions. In the \NoGSD\ run, the difference arises, because in diffuse gas the fraction of small grains is higher than in the \citetalias{1977ApJ...217..425M} GSD, leading to an underestimate of grain surface effects in this regime. At early and intermediate times none of our models reproduces the observational data in the Milky Way well. At late times, all models except \NoPr\ are in reasonable agreement with the observations.

\subsubsection{Metallicity \& Grain Size Dependence}

\begin{figure}
\includegraphics[width=0.45\textwidth]{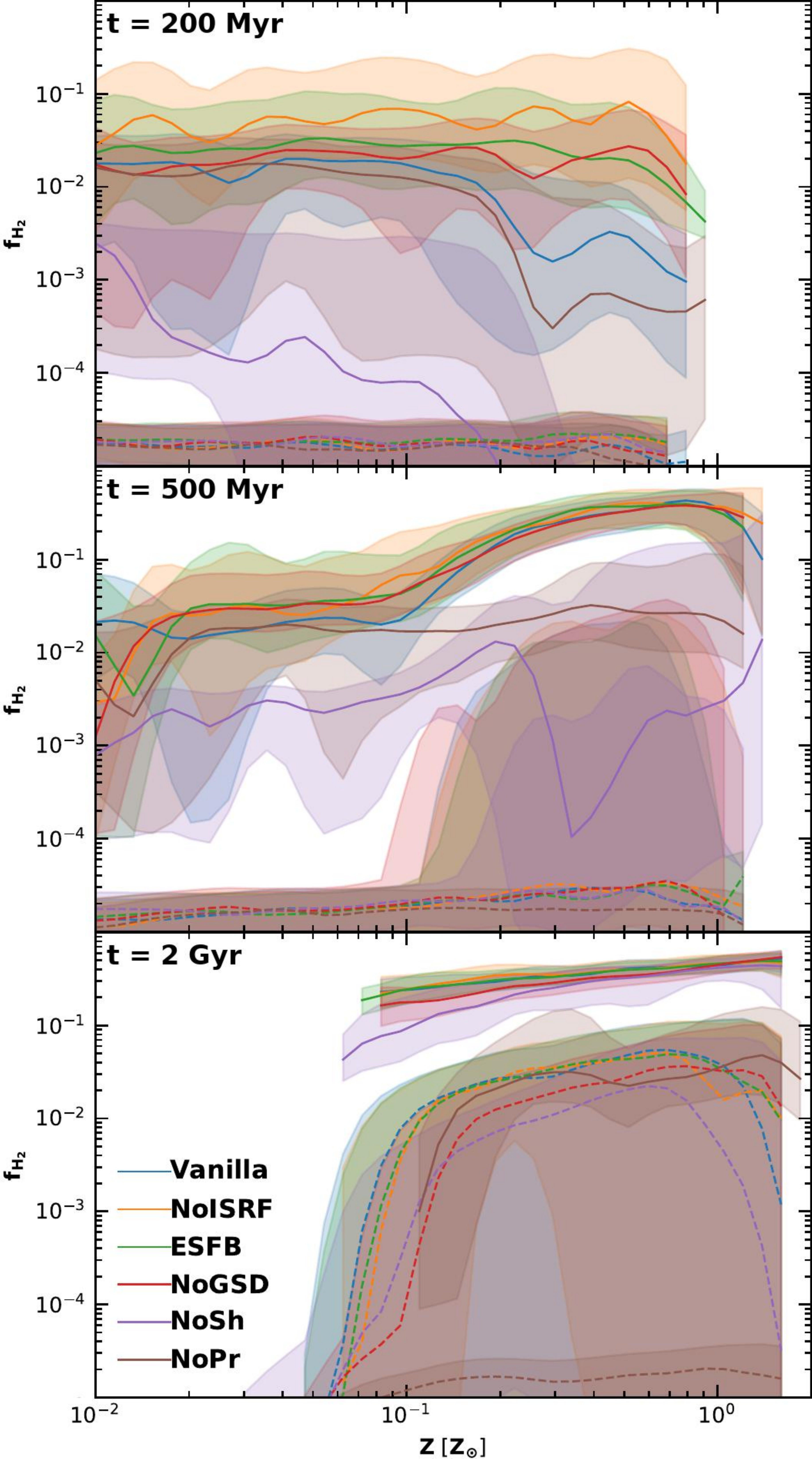}
\caption{The molecular fraction as a function of metallicity at early (top), intermediate (center) and late (bottom) times for the different runs. We show the median relation in dense gas ($n_{\text{H}} > 1\,\text{cm}^{-3}\,$; solid lines) and in diffuse gas ($n_{\text{H}} < 0.5\,\text{cm}^{-3}\,$; dashed lines). The shaded area shows the range of values between the 25th and 75th percentile of each relation.}\label{fig:H2vsZ}
\end{figure}

\begin{figure}
\includegraphics[width=0.45\textwidth]{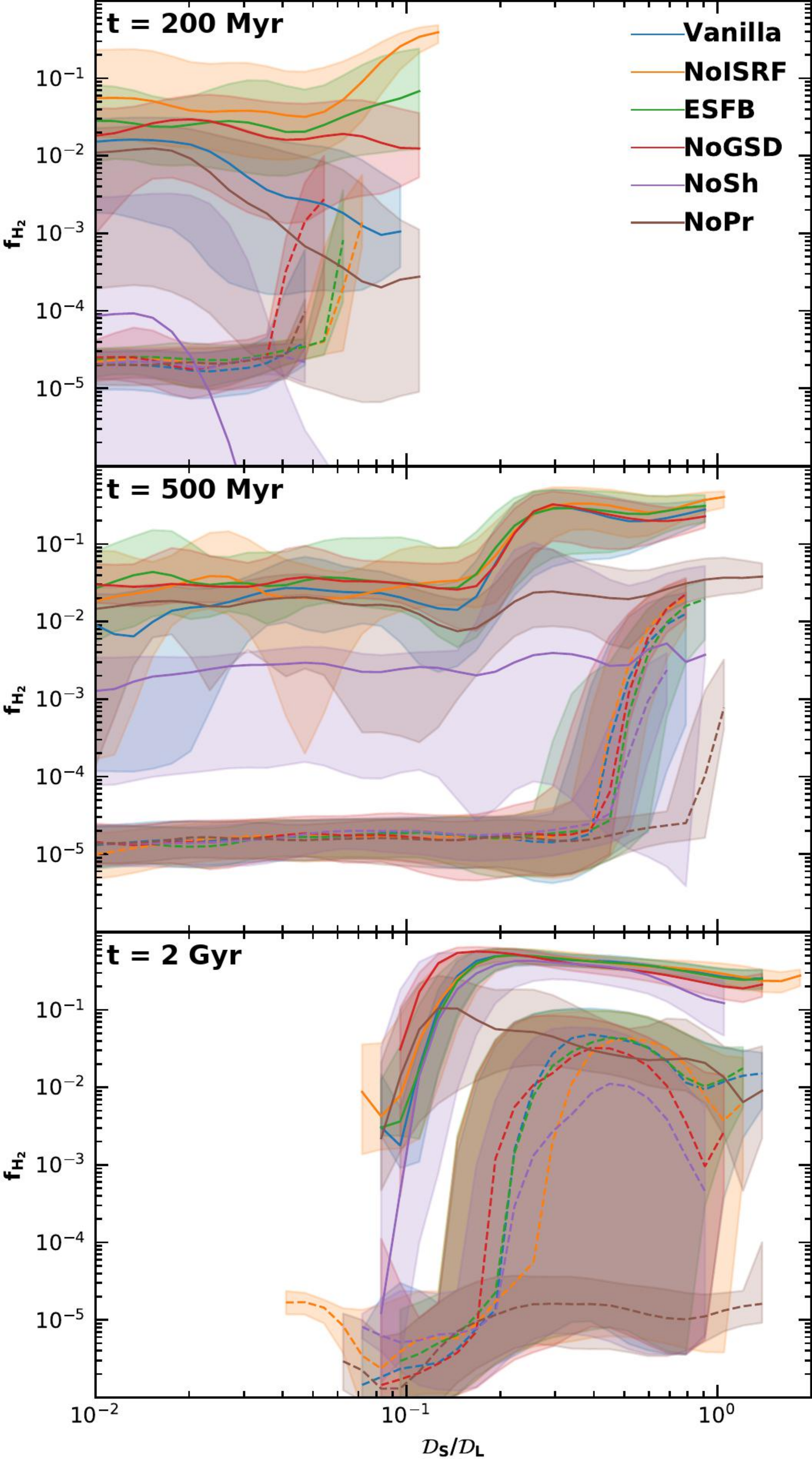}
\caption{The molecular fraction as a function of the small-to-large-grain ratio at early (top), intermediate (center) and late (bottom) times for the different runs. We show the median relation in dense gas ($n_{\text{H}} > 1\,\text{cm}^{-3}\,$; solid lines) and in diffuse gas ($n_{\text{H}} < 0.5\,\text{cm}^{-3}\,$; dashed lines). The shaded area shows the range of values between the 25th and 75th percentile of each relation.}\label{fig:H2vsDsDl}
\end{figure}

\begin{figure}
\includegraphics[width=0.45\textwidth]{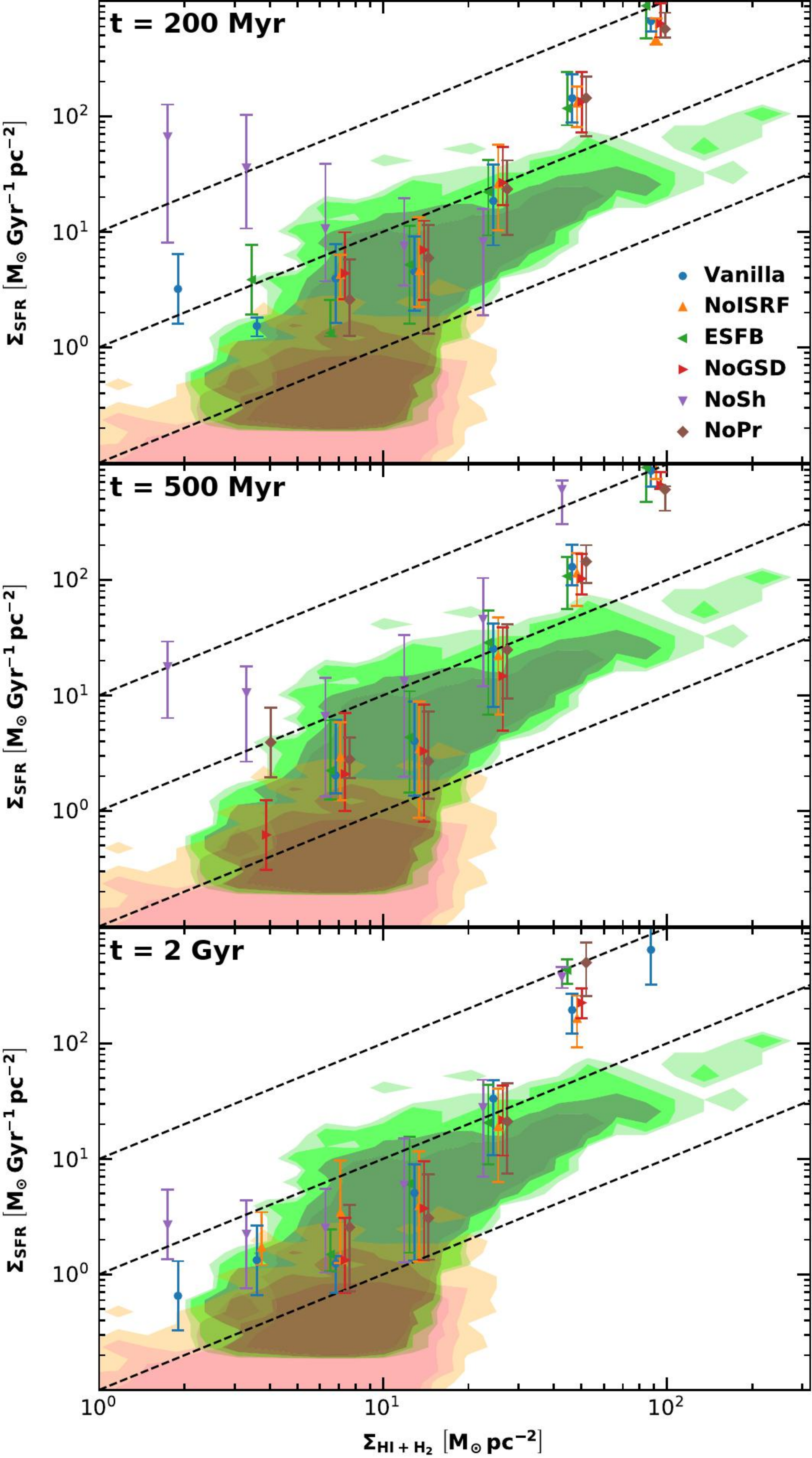}
\caption{The neutral gas KS relation at early (top), intermediate (center) and late (bottom) times for the different runs. The three dashed lines in each panel correspond to gas depletion times $\tau_{\text{depl}} = \Sigma_{\text{HI + H}_{2}}/\Sigma_{\text{SFR}}$ from top to bottom of 100 Myr, 1 Gyr and 10 Gyr, respectively. We are comparing the simulation results, which are plotted as errorbars, to the same observational sample as \citet{2019MNRAS.484.1687L}, which is taken from \citet{2008AJ....136.2846B, 2010AJ....140.1194B}.}\label{fig:KS_Hn}
\end{figure}

\begin{figure}
\includegraphics[width=0.465\textwidth]{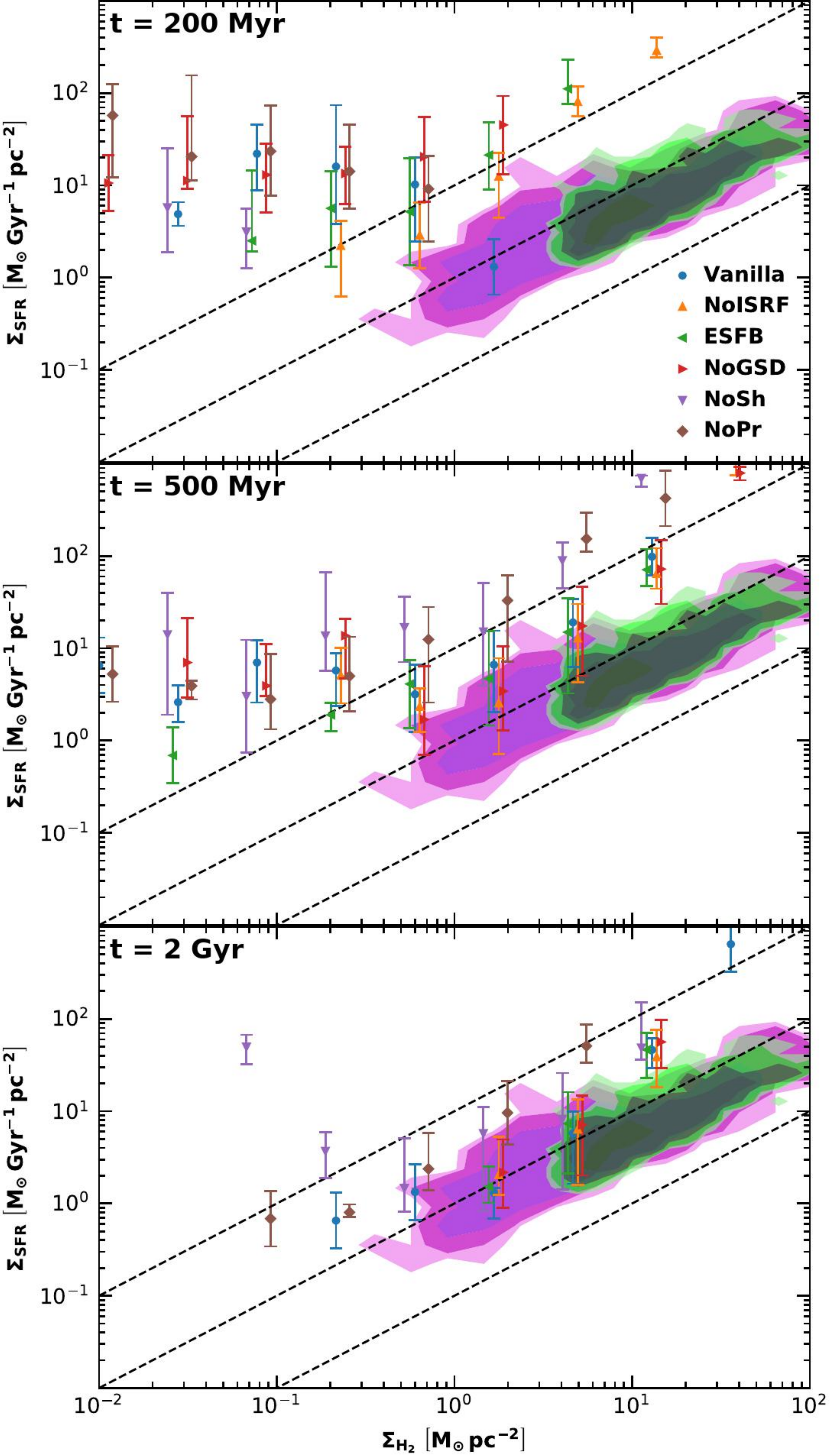}
\caption{Same as Figure~\ref{fig:KS_Hn}, but for the molecular KS relation. Observational data plotted as contours are taken from \citet{2008AJ....136.2846B} and \citet{2011AJ....142...37S} and are the same as in \citet{2019MNRAS.484.1687L} as well.}\label{fig:KS_H2}
\end{figure}

In Figure~\ref{fig:H2vsZ}, we show the molecular fraction as a function of metallicity in order to verify if our model is consistent with previous work (\citetalias{2017MNRAS.467..699H}, \citetalias{2018MNRAS.474.1545C}) and examine further differences between the models. In \citetalias{2018MNRAS.474.1545C}, the relation in the diffuse gas ($n_{\text{H}} < 10\,\text{cm}^{-3}\,$) is flat with low values of \fmol at early times. At later times, there are some diffuse particles with relatively high molecular fraction at high $Z$. In the dense gas ($n_{\text{H}} > 10\,\text{cm}^{-3}\,$), the molecular fraction increases with increasing metallicity. Due to our different subgrid prescription, where particles already start hosting dense clouds at lower densities, we consider particles to be `dense' if their densities are above $n_{\text{H}} = 1\,\text{cm}^{-3}\,$ and consider them to be diffuse if their densities are below $n_{\text{H}} = 0.5\,\text{cm}^{-3}\,$. We chose these intervals to be seperated, in order to increase the contrast between the relations. 

In all of our simulations, the molecular fraction is flat in the diffuse gas at early and intermediate times, taking values of \fmol$ \sim 2\times 10^{-5}\,$. This value is much larger than the one in \citetalias{2018MNRAS.474.1545C}, because they neglect formation of H$_2$ in the gas phase. We note that the uniform value of \fmol$ \sim 2\times 10^{-5}\,$ with metallicity in the diffuse ISM arises, because the reaction rate coefficients only depend on the temperature, which is almost constant at $T_\text{diff} \lesssim 10^4\,\text{K}\,$ in the diffuse ISM. The molecular fraction in the diffuse ISM is therefore expected to be close to the equilibrium value at this temperature independent of metallicity.

In the dense gas, the relation at early times is flat at around \fmol$ \sim 1 - 4 \,\%\,$ in all runs, except in \NoSh\ where the molecular fraction decreases as metallicity increases. The molecular fraction is higher in the run without ISRF, as there is less photodissociation. There is no positive correlation with metallicity yet, because diffusion initially delays dust growth \citepalias{2022arXiv220205243R}. 

At intermediate times, in the runs with self-shielding and H$_2$ formation on grains, the molecular fraction in the dense gas exhibits a positive correlation with the metallicity at $Z \gtrsim 0.1\, Z_{\sun}\,$. Below this metallicity, gas phase processes in combination with self-shielding dominate, keeping the molecular fraction slightly above the percent level. In the run without formation on grains, the relation is flat. In particular there is no dip at large Z indicating that shielding due to dust extinction efficiently attenuates the ISRF in this regime. In the run without self-shielding, the relation is also flat, but with typical values, that are an order of magnitude lower at around \fmol$ \sim 0.1 \,\%\,$. There is a dip at $Z \sim 0.3 \,Z_{\sun}\,$, probably since gas particles with higher metallicities are closer to stars and thus are subject to stronger feedback. However, at larger metallicities the molecular fraction increases, indicating that extinction and formation of H$_2$ on grain surfaces counteract the strong photodissociation in this regime.

At late times, there is a mildly positive correlation between the molecular fraction and metallicity for the dense gas in all runs with formation on grain surfaces. Molecular fractions increase here from \fmol$ = 20\,\%\,$ at $Z = 0.1\,Z_{\sun}\,$ to \fmol$ = 60\,\%\,$ at $Z \gtrsim 1\,Z_{\sun}\,$. The value in \NoSh\ at low $Z$ is slightly lower than in the other runs with molecule formation on dust. In \NoPr\ the molecular fraction in dense gas does not significantly exceed the percent level, even at high $Z$, while it exhibits the same molecule poor flat relation as at early and intermediate times in the diffuse gas. In the other runs, the metal rich ($Z \gtrsim 0.1 \,Z_{\sun}$), diffuse gas becomes enriched with molecules at the percent level at late times. Through comparison with the \NoPr\ run it becomes apparent that the process responsible for this is formation on dust grains. This is a quite plausible scenario, as high small-to-large-grain ratios, which drive dust growth and formation of molecules, are typically encountered in the diffuse ISM. Since there are only small differences between \NoGSD\ and the other runs, it further becomes apparent that here the shape of the GSD is only subdominant. However, even in \NoGSD\ grain growth is driven by small grains and thus the GSD may implicitly affect the molecular fraction here.

In order to further study the effect of grain size, in Figure~\ref{fig:H2vsDsDl} we show the molecular fraction as a function of the small-to-large-grain ratio \DsDl, which is defined to be the mass ratio of large and small grains, where we set the grain radius separating the two regimes to be at $a = 0.03\,\micron\,$. With this definition the value of \DsDl corresponding to a \citetalias{1977ApJ...217..425M} GSD with maximum grain radii $a = (0.25 - 1)\,\micron\,$ is roughly \DsDl$\sim 0.1 - 0.2$.

At all times, the molecular fraction is constant taking a low value below a certain threshold value of \DsDl and then jumps to a significantly higher roughly constant value at high \DsDl. The threshold value is higher in the diffuse gas and is different at different times. At early times there is no jump at high \DsDl in the dense gas, indicating that dust related molecular physics are still rather irrelevant at this stage. Nonetheless, there is a steep rise in \fmol in the diffuse gas at \DsDl$\sim 3-4 \times 10^{-2}\,$. At intermediate times the threshold value of \DsDl in the dense gas is at around \DsDl $\sim 0.2 - 0.3$. At this point the molecular fraction jumps from \fmol$\sim 2\,\%\,$ to \fmol$\sim 20\,\%\,$ in the runs with self-shielding and formation on grain surfaces. In the runs without self-shielding or formation on grain surfaces no significant jumps are seen in the dense gas relations. In the diffuse gas the threshold value of \DsDl is at around \DsDl$\sim 0.4 - 0.5$. Here the molecular fraction jumps by about 3 orders of magnitude from \fmol$\sim 10^{-5}$ to \fmol$\sim 0.01$ in all runs except in the run without formation on grain surfaces, indicating that the process responsible for this jump is the formation of molecules on grain surfaces. At late times the threshold values have shifted to smaller values. In the dense gas the threshold value is now closer to \DsDl$\sim 0.1$. Despite there being a slight jump in \NoPr\, the maximum achieved values of \fmol do not significantly exceed the percent level. In the diffuse gas the threshold is now at \DsDl$\sim 0.2$. The \NoPr\ run does not exhibit a jump in the diffuse gas.

\subsection{Star Formation Law}

Observations have made it clear that there is a relation between the SFR and gas surface densities \citep[e.g.][]{1959ApJ...129..243S, 1998ApJ...498..541K, 2002ApJ...569..157W, 2008AJ....136.2782L}. This relation, which is often referred to as the Kennicutt--Schmidt (KS) relation, has been used as a benchmark to calibrate star formation and feedback models \citep{2016ApJ...833..202K, 2019MNRAS.484.2632S} and as a basis for star formation models \citep{2008MNRAS.387.1431D, 2012MNRAS.426..140D}. Moreover, it has been shown that an even more tight correlation can be found between the SFR and molecular gas column densities, while there is almost no correlation between the SFR and the H\,\textsc{i} column density \citep{2008AJ....136.2846B}. The relation between the SFR and the molecular gas column densities, which we refer to as the molecular KS relation, may be used as a benchmark for chemical evolution models as any model that tries to simulate galaxies with star formation and chemical evolution should reproduce the KS relation both with respect to the neutral (H \textsc{i} and H$_2$) and the molecular gas.

In Figures~\ref{fig:KS_Hn} and \ref{fig:KS_H2}, we compare the resulting KS relations in our simulations to the observations of \citet{2008AJ....136.2846B, 2010AJ....140.1194B} (green and red contours) and \citet{2011AJ....142...37S} (purple contours). The different contour levels are corresponding to the number density of data points within each bin, with darker colors corresponding to more data points. We compute the column densities of H\,\textsc{i}, H$_2$ and the SFR by projecting them onto a grid as described above in Section \ref{sec:H2vsn}. We then exclude all grid cells with vanishing $\Sigma_{\text{SFR}}$ and assign the remaining cells to 10 logarithmically spaced bins on the $\Sigma_{\text{HI+H}_{2}}$ and the $\Sigma_{\text{H}_{2}}$ axis to compute the median and 25th (75th) percentile in each bin. The results of this procedure are shown as error bars in Figures~\ref{fig:KS_Hn} and \ref{fig:KS_H2}. 
The neutral gas KS relation in Figure~\ref{fig:KS_Hn} hardly evolves in time in all models, except in \NoSh, where at early times, we obtain high SFR column densities at low neutral gas column densities. This happens, because without self-shielding, the gas in star-forming region gets ionized, reducing the neutral gas column density. In the other runs, we get a similar neutral KS relation, which agrees rather well with the observations, but overshoots at very high and very low column densities. This is in good agreement with the results from \citet{2019MNRAS.484.2632S}, which this work is based on. \citet{2022arXiv220100970O} used high resultion simulations of superbubbles to study the dependency of the shock momentum on the properties of the ISM, like metallicity and density. Based on their calculations they modified the {\sc Gadget3-Osaka} model and performed simulations of an isolated Milky Way-like galaxy.
They report a neutral KS-law which agrees well with the observations also at relatively low gas column densities of $\text{log}\,\Sigma_{\text{HI + H}_2}\left[\text{M}_{\sun}\,\text{pc}^{-2}\right] \sim 0.5$. In this regime our simulations tend to slightly overpredict the column density. This indicates that the details of the feedback prescription can affect the resulting star formation law in non-trivial ways. At large column densities of $\log\Sigma_{\text{HI} + \text{H}_2}~[\text{M}_{\sun}\,\text{pc}^{-2}] > 1.5$, both our model and the one of \citet{2022arXiv220100970O} overshoot considerably, while the model of \citet{2019MNRAS.484.1687L} predicts values that are closer to the observations. \citet{2019MNRAS.484.1687L} uses a star formation prescription, in which the star formation efficiency $\epsilon_{\rm SF}$ is regulated by turbulence. At high column densities, we would indeed expect the gas to be stabilized by supersonic turbulence, an effect that is not captured by our model.

The molecular KS relation shown in Figure~\ref{fig:KS_H2} evolves in time, and there are notable differences between the models. At early times molecular depletion times $\tau_{\text{depl}}^{\text{mol}} = \Sigma_{\text{mol}}/\Sigma_{\text{SFR}}$ are of the order of $100\,\text{Myr}\,$ in the run without ISRF, but are significantly shorter at low molecular gas column densities, in all other runs. In the \ESFB\ run, the depletion time is slightly higher at low molecular column densities, but is roughly on the order of $100 \, \text{Myr}\,$ at intermediate and high molecular column densities. In the other runs, the relation between $\Sigma_{\text{SFR}}$ and $\Sigma_{\text{mol}}$ is initially almost flat. This is because initially, there is not enough dust to maintain the H$_2$ in the star-forming gas, which is predominantly dissociated, leading to low molecular column densities at high SFR column densities. At intermediate times, as the ISM becomes sufficiently enriched with metals and dust and the star formation activity slows down, more molecular hydrogen can survive and molecular depletion times increase. At this time self-shielding still plays an important role, especially in gas which is enriched with higher levels of H$_2$ due to formation on grain surfaces. In the \NoSh\ and \NoPr\ runs, depletion times of $\tau_{\text{depl}}^{\text{mol}} \sim 100\,\text{Myr}\,$ are achieved, in the former due to increased production of H$_2$ on grain surfaces and in the latter due to self-shielding, dust extinction and the overall weaker ISRF. In all other runs, slightly longer depletion times ranging from $500\,\text{Myr}\,$ to $1\,\text{Gyr}\,$ are achieved at intermediate molecular column densities $\Sigma_{\text{mol}}\sim 1 - 10 \text{M}_{\sun}\,\text{pc}^{-2}\,$. At low and high molecular column densities all models exhibit shorter depletion times, i.e. tend to overshoot a linear molecular KS relation, with fixed depletion time. The overshooting at high $\Sigma_{\text{mol}}$ is related to the overshooting in the neutral KS-relation and probably related to the SF and feedback prescription.  The low molecular column densities at relatively high SFR column densities, are symptomatic of the dust growth which is still in progress, leaving regions which are dense enough for star formation, but not yet enriched with enough dust to keep the molecular hydrogen from dissociation in the ISRF. At later times the depletion times are even further reduced in all runs and are now in rather good agreement with the observed molecular KS law with a constant depletion time of roughly $1 \,\text{Gyr}\,$ over a wide range of molecular column densities. There is still some overshooting in the relation at large and small molecular column densities, which is also observed in the neutral KS relation, indicating that it is related to the star formation and feedback model and not to the chemical evolution model.

\section{Discussion and Conclusions}\label{sec:discussion}

We presented here the results of our fully self-consistent treatment of molecular hydrogen enrichment with the GSD in a simulation of an isolated Milky-Way-like galaxy. We adapted the model for hydrogen enrichment by \citetalias{2018MNRAS.474.1545C} to our full treatment of the GSD with diffusion of dust and metals, which is based on the works of \citetalias{2019MNRAS.482.2555H}, \citetalias{2020MNRAS.491.3844A} and \citetalias{2022arXiv220205243R}. Most of the details in the numerical treatment are in line with \citetalias{2018MNRAS.474.1545C}, but there are some important differences. We follow the non-equilibrium evolution of all species in the chemistry network self-consistently with the hydrodynamic evolution of the gas and take into account the effect of gas phase production and dissociation of molecular hydrogen. Furthermore, instead of using an inverse square law to propagate the ISRF from star-forming particles, we simply use an SPH kernel estimate to smooth the star formation rate column density field, due to its straightforward implementation into SPH codes. We also allow the fraction of dense clouds, which determines the relative importance of the subgrid physics to grow with density, rather than using a fixed value above a density threshold as discussed in \citetalias{2022arXiv220205243R}. We compared our full model to a number of reduced or modified version of the model, in order to understand the regimes in which each process is important.
\begin{enumerate}
    \item In order to adequately model the molecular abundance during epochs of strong star formation, a proper treatment of the ISRF is needed. Event-based energy injection like in our \ESFB\ model tends to be significantly less efficient than continuous energy injection via a smoothed interstellar radiation field stored by the gas particles. This is because timesteps between energy injection events would need to be comparable to the short timesteps involved in chemistry calculations in order to leave a lasting imprint on the chemical abundances. This criterion on the other hand can be circumvented by informing the chemistry network of the presence of heating due to an ISRF in the continuous model.
    \item In low metallicities, self-shielding and gas phase processes are critical to obtain reasonably high molecular abundances. Neglecting self-shielding will delay molecular enrichment lead to overall less molecular gas.
    \item Formation of molecules on the surfaces of dust grains is the main driver of molecular enrichment. In dense gas, inclusion of this process can increase molecular abundances by more than an order of magnitude.
    \item The effects of grain size on the molecular abundance are mostly minor, but there are regimes where notable differences between \NoGSD\ and the other runs arise. In particular in regimes which are similar to the physical conditions at early times, where the GSD is dominated by large grains which tend to be rather unimportant for grain surface physics, the treatment which is agnostic about the GSD tends to overestimate the effect of dust. On the other hand, in regimes where the physical conditions are resembling the Milky Way at late times, the effect of small grains in the diffuse ISM is underestimated, leading to overall less molecular gas in this regime. However, due to the nature of our idealized ICs, deciding whether or not these regimes are indeed realized in nature is out of the scope of this study.
    \item Our \Vanilla\ model converges towards a solution at late times that is consistent with local spiral galaxies. In particular, we reproduce the relation between molecular fraction and column density in the Milky Way \citep{2008ApJ...680..384W, 2006ApJ...636..891G} and the neutral and molecular KS law in nearby spiral galaxies \citep{2008AJ....136.2846B, 2010AJ....140.1194B, 2011AJ....142...37S}. Furthermore the final value of the global molecular gas fraction in the disk is consistent with the observed value of \fmol$\sim 20\,\%\,$ in galaxies with masses similar to the Milky Way galaxy \citep{2018MNRAS.476..875C}. It is however important to note, that this result might simply be due to our choice of the parameter $\alpha$, which has been chosen to give a global fraction of dense gas in rough agreement with the galactic molecular fraction. We stress that the molecular KS law in our simulation depends on the evolutionary stage of the galaxy and the GSD in our simulation. This is a general feature of this model, which can be tested with observations of galaxies with low metallicites and/or strong star formation activity.
    \item At high and low surface densities our models tend to overestimate the star formation rate surface density. This might be symptomatic of the feedback and star formation prescription as is indicated by \citet{2019MNRAS.484.1687L} and \citet{2022arXiv220100970O}, who report better agreement with the observations at lower column densities in their models.
\end{enumerate}

As stressed by \citetalias{2018MNRAS.474.1545C}, the results of our calculations are based on a subgrid treatment of unresolved scales and therefore depend on a number of model parameters as well as some questionable assumptions. Apart from a few exceptions, cooling and chemistry are calculated for gas particles solely based on the SPH density, temperature and metallicity, but do not take into account the multiphase nature of the gas on unresolved scales. This kind of treatment might have led to an overestimate of molecule dissociation. Furthermore as mentioned above, the immediate impact SN Feedback has on the chemical abundances and ionization state in the shocked gas has been completely neglected in our feedback model. In models where cooling is disabled for the shocked gas, the chemical abundances are locked in the state before the energy injection. Once the gas is allowed to cool again, the chemical abundances and thus also the cooling rates will be inconsistent, with the expectation of a fully ionized plasma. As shown by \citet{1977ApJ...216..713K}, molecules would be fully dissociated by such strong shocks. Not accounting for the ionization of the gas and the dissociation of molecules in SN shocks, can thus lead to wrong predictions.
A suitable treatment of these aspects in future simulations might lead to further insights. Another serious limitation is the treatment of the dense clouds on unresolved scales. In our model we have kept their densities and temperatures fixed and assumed, that their mass fraction grows linearly with the density until it saturates at unity. In reality, dense clouds come with a wide range of densities and temperatures.
Thus, it might be fruitful to draw pairs of cloud densities and temperatures from a probability distribution, which agrees with observations or to perform averages over densities and temperatures.
The assumptions on subgrid structures could be further constrained with observations by cross-correlating observations on large and small scales.

Finally, the results from a simulation of an isolated galaxy are inherently limited by all assumptions that go into the initial conditions and the employed additional physics modules. In our simulation metallicities and elemental abundances are set to primordial values despite the fact that the galaxy already has an established stellar disk and bulge, making up most of its baryonic mass. Furthermore we do not include AGN feedback, and by design environmental effects cannot be taken into account. It is therefore important to eventually test our models in cosmological simulations of structure formation, which alleviaties some of the limitations of isolated-galaxy simulations.

In conclusion, we have developed a self-consistent non-equilibrium subgrid model for the evolution of molecular hydrogen in the ISM under the influence of an ISRF and dust grains, which is able to reproduce a number of observational constraints from the Milky Way galaxy and local spirals. We have identified which processes are most important under which physical conditions. In our star formation model, where the star formation rate does not explicitly depend on the molecular gas, the strong correlation referred to as the molecular KS law is only reproduced under very specific physical conditions which are typical for the nearby spiral galaxies in which it is observed. We therefore predict that in systems with low metallicities and strong star formation the correlation between molecular gas and star formation would be weaker. 

\section*{Acknowledgements}

We thank the referee for his insightful comments and suggestions that helped to improve the quality of this communication. We thank Alessandro Lupi for kindly providing us with the observational data and the script that went into producing the contours in Figures~\ref{fig:KS_Hn} and \ref{fig:KS_H2}.
Our numerical simulations and analyses were carried out on our local cluster {\sc Orion} at Osaka University.
This work was partly supported by the JSPS KAKENHI Grant Number JP17H01111, 19H05810, 20H00180. 
KN acknowledges the support from the Kavli IPMU, World Premier Research Center Initiative (WPI).
HH thanks the Ministry of Science and Technology (MOST) for support through grant
MOST 108-2112-M-001-007-MY3, and the Academia Sinica
for Investigator Award AS-IA-109-M02.

\section*{Data Availability}

Data related to this publication and its figures are available on request from
the corresponding author.



\bibliographystyle{mnras}
\bibliography{references_v1} 

\clearpage
\appendix

\section{Dependence on the Initial Conditions}\label{Appendix:ICs}

\begin{figure}
\includegraphics[width=0.45\textwidth]{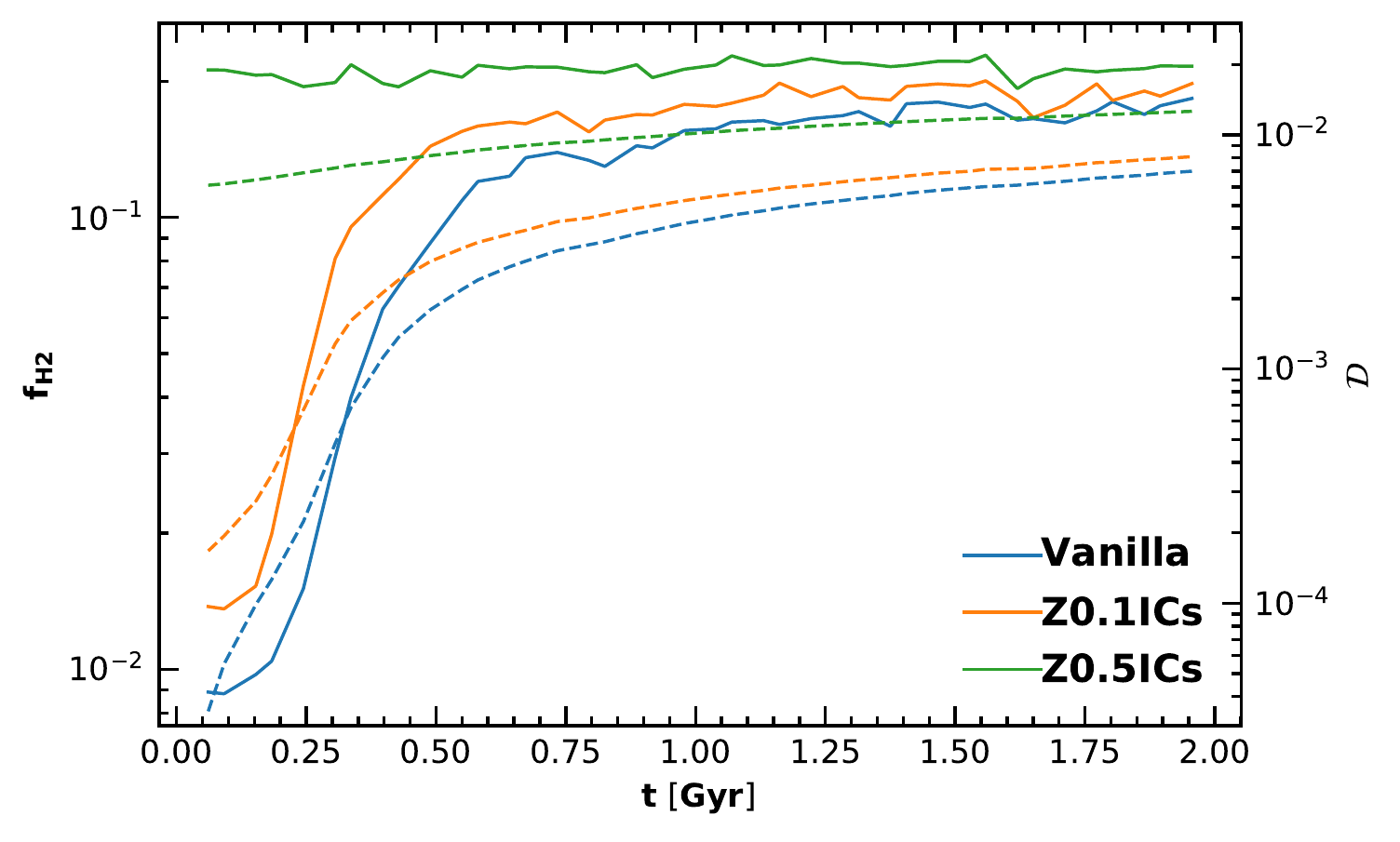}
\caption{Same as fig. \ref{fig:H2history} for the different ICs. Solid lines correspond to the molecular fraction (left y-axis) and dashed lines to the dust-to-gas ratio (right y-axis).}\label{App:fig:H2history}
\end{figure}

In order to verify
the robustness of our results,
we have run two additional simulations with our fiducial \Vanilla\ model with initial conditions, in which the gas in the disk is already enriched with metals and dust. In one set of initial conditions, in the following referred to as \Early, we adopt a metallicity of $Z = 0.1 Z_{\sun}$ everywhere in the disk, and a GSD following directly the yield relation (log-normal GSD with\DtoZ $= 0.1$). In the other set of initial conditions, in the following referred to as \Late, we adopt a metallicity of $Z = 0.5 Z_{\sun}$ and a GSD resembling the MRN GSD for graphite in the range 0.05--1\,$\micron\,$ with an extreme dust-to-metal ratio of \DtoZ $= 1$. In all runs, the Hydrogen and Helium is initially fully atomic everywhere.

Figure~\ref{App:fig:H2history} shows the global molecular fraction and dust-to-gas ratio in the disk as a function of time for the different ICs. In the \Late\ run, the molecular fraction reaches $\sim 20 \%$ already in the first snapshot, i.e. after $50\,\text{Myr}\,$, indicating, that the timescales are short enough for the molecular fraction to be well described by the equilibrium of the formation on grain surfaces and the attenuated ISRF. The molecular fraction in the runs with initially already enriched runs is higer at all runs, with more enrichment, but the differences become smaller at later times, when the GSD has mostly converged and the metallicity only increases slowly.

\begin{figure}
\includegraphics[width=0.45\textwidth]{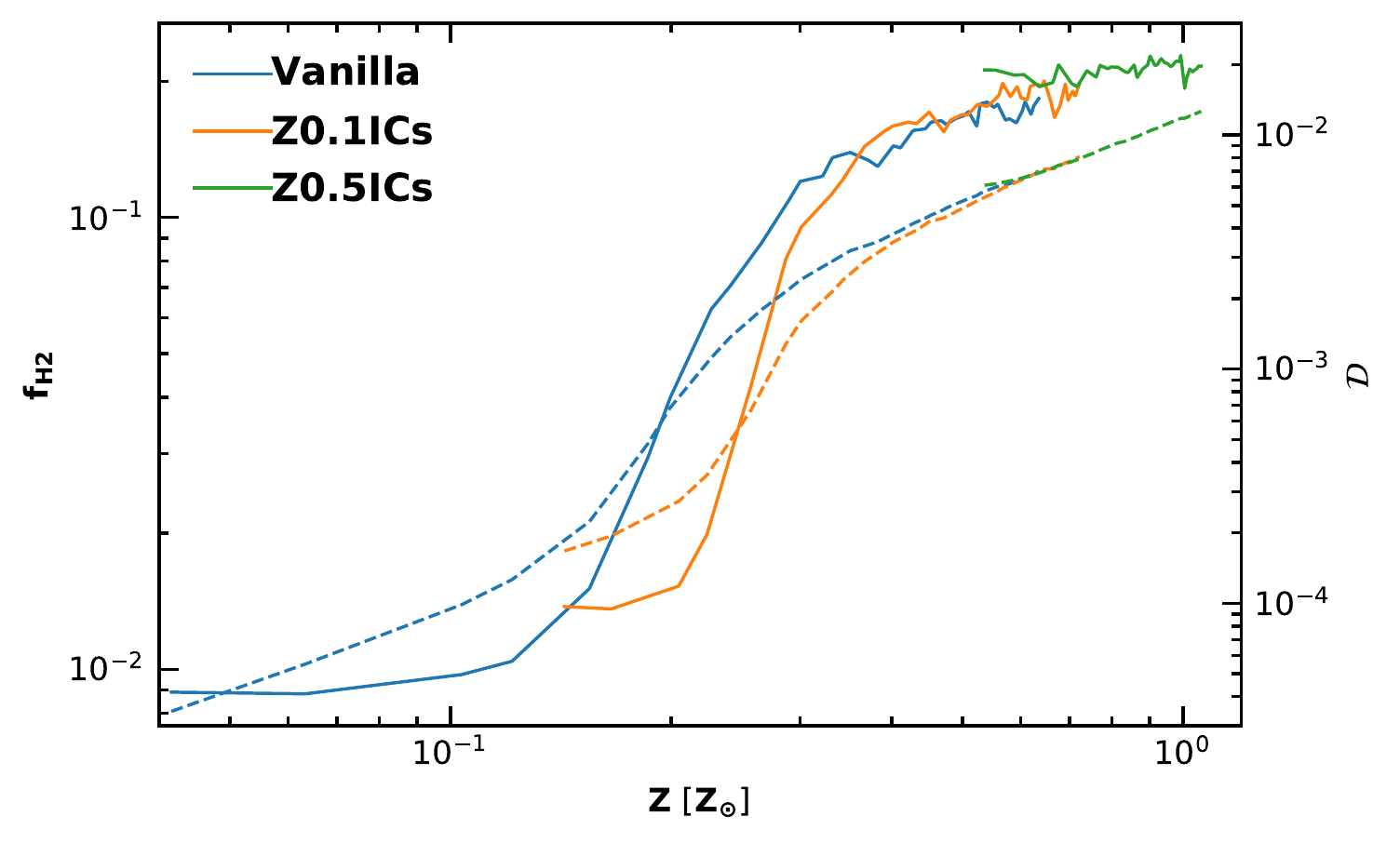}
\caption{Same as fig. \ref{App:fig:H2history} but with metallicity as new time variable.}\label{App:fig:H2vsZ}
\end{figure}

Since the initial conditions could roughly be understood as slightly shifted in time, it might be better to use the metallicity as time-variable instead of the time. Arguably, it can then be claimed that our results for a quantitiy are robust with respect to the ICs, if its trend with metallicity is consistently extended to higher $Z$. To this end, in Figure~\ref{App:fig:H2vsZ} we show the molecular fraction and the dust-to-gas-ratio as a function of global metallicity for the different ICs.

As can be seen, the dust-to-gas ratio in the \Early\ run follows a slightly different track than the \Vanilla\ run at $Z \sim (0.1 - 0.2) Z_{\sun}$, where \Dtot starts to increase already at lower $Z$ in the \Vanilla\ run. At high $Z$ all runs converge and the runs with higher initial $Z$ extend the relation plausibly to higher $Z$. This is likely at the heart of the differences in the molecular fraction, which is lower in the \Early\ run at the metallicities where \Dtot is lower as well, but also converges at higher $Z$.
The differences at low $Z$ might be due to the different history of feedback, which would keep the GSD closer to the yield distribution at early times and due to a different dilution phase, since in the \Early\ run all parts of the disk are already enriched with dust from the beginning, whereas in the \Vanilla\ run dust growth is initially suppressed while the metals spread through the disk.

From Figures~\ref{App:fig:H2history} and \ref{App:fig:H2vsZ} we can see that the GSD and the molecular fraction have roughly converged for the different ICs at the point, where $f _{\text{H}_2}\gtrsim 0.1$, which is around $t \gtrsim 500 \,\text{Myr}\,$ in the \Vanilla\ run. Thus we can be confident that our results at intermediate times and late times are robust, while the results at early times should be checked with initial conditions consistent with cosmological models.

%

\bsp	
\label{lastpage}
\end{document}